\documentclass[fleqn,10pt]{SelfArx} 

\definecolor{color1}{RGB}{0,0,90} 
\definecolor{color2}{RGB}{0,20,20} 

\usepackage{hyperref} 
\hypersetup{hidelinks,colorlinks,breaklinks=true,urlcolor=color2,citecolor=color1,linkcolor=color1,bookmarksopen=false,pdftitle={Title},pdfauthor={Author},urlcolor=blue}

\usepackage{graphicx}
\usepackage[square]{natbib}
\usepackage{amsmath}
\usepackage{multirow}
\usepackage{subfigure}
\usepackage{multirow}

\newcommand{\pd}{\partial}

\newcommand{\n}[1]{\mathrm{#1}}
\newcommand{\parde}[2]{\frac{\pd #1}{\pd #2}}

\JournalInfo{Published in International Journal of Refrigeration, Vol. 63 , 48-62, 2016} 
\Archive{\href{http://dx.doi.org/10.1016/j.ijrefrig.2015.08.022}{DOI: 10.1016/j.ijrefrig.2015.08.022}} 

\PaperTitle{The lifetime cost of a magnetic refrigerator} 

\Authors{R. Bj\o{}rk, C.R.H. Bahl and K. K. Nielsen} 
\affiliation{\textit{Department of Energy Conversion and Storage, Technical University of Denmark - DTU, Frederiksborgvej 399, DK-4000 Roskilde, Denmark}} 
\affiliation{*\textbf{Corresponding author}: rabj@dtu.dk} 

\newcommand{\keywordname}{Keywords} 

\Abstract{The total cost of a 25 W average load magnetic refrigerator using commercial grade Gd is calculated using a numerical model. The price of magnetocaloric material, magnet material and cost of operation are considered, and all influence the total cost. The lowest combined total cost with a device lifetime of 15 years is found to be in the range \$150-\$400 depending on the price of the magnetocaloric and magnet material. The cost of the magnet is largest, followed closely by the cost of operation, while the cost of the magnetocaloric material is almost negligible. For the lowest cost device, the optimal magnetic field is about 1.4 T, the particle size is 0.23 mm, the length of the regenerator is 40-50 mm and the utilization is about 0.2, for all device lifetimes and material and magnet prices, while the operating frequency vary as function of device lifetime. The considered performance characteristics are based on the performance of a conventional A$^{+++}$ refrigeration unit. In a rough life time cost comparison between the magnetic refrigeration device and such a unit we find similar costs, the former being slightly cheaper, assuming the cost of the magnet can be recuperated at end of life.}

\begin{document}

\flushbottom 

\maketitle 


\thispagestyle{empty} 

\section{Introduction}
Magnetic refrigeration is a promising efficient and environmentally friendly technology based on the magnetocaloric effect. A substantial number of scientific magnetic refrigeration devices have been published \citep{Yu_2010,Kitanovski_2015}, but so far the technology has yet to be commercialized. The main challenge for this is the relatively small magnetocaloric effect present in currently used magnetocaloric materials; the benchmark magnetocaloric material, Gd, has an adiabatic temperature change of less than 4 K in a magnetic field of 1 T \citep{Dankov_1998,Bjoerk_2010d}, depending on purity. Therefore a regenerative process, called active magnetic regeneration (AMR), is used to produce the desired temperature span \citep{Barclay_1982}.

An important aspect in the commercialization of magnetic refrigeration is proving the often mentioned (potentially) high efficiency of magnetic refrigeration devices. Furthermore, it is crucial to show that these devices will have a lower lifetime cost than vapour compression based devices. Magnetic refrigeration devices will have a larger construction cost than vapour compression devices, due to the permanent magnet material needed to provide the magnetic field in the device. However, if the operating cost is significantly lower than compression based devices, then magnetic refrigeration devices may be overall cheaper. Determining the operation and construction cost of a magnetic refrigeration device is the purpose of this paper.

The total construction cost of a magnetic refrigeration unit has previously been considered by a number of authors. \citet{Rowe_2011} defined a general performance metric for active magnetic regenerators, which included the cost and effectiveness of the magnet design as a linear function of the volume of the magnet and the generated field. A figure of merit used to evaluate the efficiency of the magnet design was introduced in \citet{Bjoerk_2008} not taking the performance of the actual AMR system into account. The optimal AMR system design, i.e. ignoring the magnet, has been considered by \citet{Tusek_2013b}.

The building cost of a magnetic refrigeration device was considered by \cite{Bjoerk_2011d}, for a device with a given temperature span and cooling power calculated using a numerical model. Both a Halbach cylinder and a ``perfect'' magnet were considered, as well as both parallel plates and packed sphere regenerators. Assuming a cost of the magnet material of \$40 per kg and of the magnetocaloric material of \$20 per kg the cheapest packed sphere bed refrigerator with Gd that produces 50 W of continuous cooling at a temperature span of 30 K using a Halbach magnet was found to use around 0.15 kg of magnet, 0.04 kg of Gd, having a magnetic field of 1.05 T and a minimum cost of \$6. The cost is dominated by the cost of the magnet. However, this calculation assumed magnetocaloric properties as predicted by the mean field theory, which is known to overestimate magnetocaloric properties compared to commercial grade Gd \citep{Bahl_2012}. Also, the operating cost of the device was not considered.

The model presented by \citet{Tura_2014} determined the total cost and optimal geometry and operating conditions for a dual-regenerator concentric Halbach configuration using a simple analytical model of an AMR. The magnetocaloric material was taken to be ideally graded, i.e. the adiabatic temperature change was defined as a linear function of temperature throughout the AMR and with a constant specific heat equal to that of Gd at the Curie temperature. Furthermore, a single particle size of 0.3 mm was considered. Both the manufacturing and the operating costs were considered and the lowest cost device was found as a function of the desired cooling power and effectiveness of the magnetocaloric material for a fixed temperature span of 50 K. For a cooling power of 50 W the system with the lowest cost had a magnetic field of 1 T, a frequency around 4.5 Hz, a utilization of 0.35 and a COP of 2. The capital costs are around \$100 and \$40 for the magnet and the magnetocaloric material, respectively, while the cost of operation is \$0.004 h$^{-1}$.

In this paper we will consider not only the construction cost of the magnetic refrigeration device, but also the operating cost. Based on these, we will calculate the overall lowest cost of the magnetic refrigeration device based on the price of the magnet material, the price of the magnetocaloric material and the expected lifetime of the device.

\section{Required device performance}
In order to get relevant cooling performance values we chose as a benchmark for this study a refrigerator appliance in the energy class A$^{+++}$ (EU-label system), specifically a well insulated appliance with a 350 L inner volume. As vapor compression devices operate differently from magnetocaloric devices it can be hard to find a fair way of making a direct comparison between the two. Thus, the intention of this paper is to identify a magnetocaloric unit with an output performance resembling that experienced from the vapor compression unit.

According to the calculation scheme of EU-directive 1060/2010, the average electrical power consumption must not exceed 8.6 W \citep{Mrzyglod_2014}. At a coefficient of performance of about 3.2, which is the operating COP for an A$^{+++}$ appliance \citep{Mrzyglod_2014}, this is equivalent to an average cooling power of about 24 W, assuming an ambient temperature of 25$^{\circ}$C. However, door openings, loading and periods of increased ambient temperature will result in an increase in the cooling power demand. In general the compressor in the device will be dimensioned for loads well above the average, and be operated in an on/off manner at times of lower cooling power demand.

Taking the values from \citet{Mrzyglod_2014}, the magnetocaloric device considered in the following will be dimensioned to deliver a maximum cooling power of $Q_\n{high}$ = 50 W for 10\% of the time and $Q_\n{low}$ = 22 W for the remaining 90\% of the time. This gives an average cooling power of $Q_\n{av}$ = 24.8 W, close to that of the considered A$^{+++}$ appliance. Thus, the AMR must be large enough to deliver 50 W, but operate most of the time at a much lower load. This will be compared to a device continuously operating at a cooling power of 24.8 W, using a volume of cold storage to increase the cooling power at times of higher demand. Throughout we apply a temperature span in the AMRs of $\Delta{}T=30$ K. We are well aware that the span in vapor compression appliances is significantly larger than this, allowing for a significant temperature span in each of the heat exchangers of about 10 K. However, an AMR based appliance will operate differently compared to a traditional appliance. One of the fundamental requirements of a fully magnetocaloric system is a low span of about 2 K in each heat exchanger. Thus, these will have to be redesigned for such a device. This can be done through increasing the areas and heat transfer, e.g. by forced convection. So with a reduction of a few degrees at each end the chosen span will resemble that experienced in a household refrigerator.

\section{Determining the performance of an AMR device}
The regenerator in a magnetic refrigeration device consists of a porous matrix of a solid magnetocaloric material and a heat transfer fluid that can flow through the matrix while rejecting or absorbing heat. The excess heat is transferred to a hot-side heat exchanger connected to the ambient while a cooling load is absorbed at the cold end of the AMR. Typically, the porous matrix is either a packed sphere bed \citep{Okamura_2005,Tura_2009} or consists of parallel plates \citep{Zimm_2007, Bahl_2008}. It has been shown, at low operating frequency, that parallel plate regenerators can produce relevant temperature spans \citep{Bahl_2012,Tusek_2013}. But for the study conducted here a packed sphere bed regenerator is considered since this is experimentally found to have superior performance compared to parallel plate regenerators \citep{Tura2011,Tura2012}. The reason for this is likely more of a practical nature rather than theoretical since it has proven very difficult to manufacture parallel plate regenerators with sufficient accuracy to meet the performance of packed spheres \citep{Nielsen2013a}.

\begin{figure}[!t]
  \centering
  \includegraphics[width=1\columnwidth]{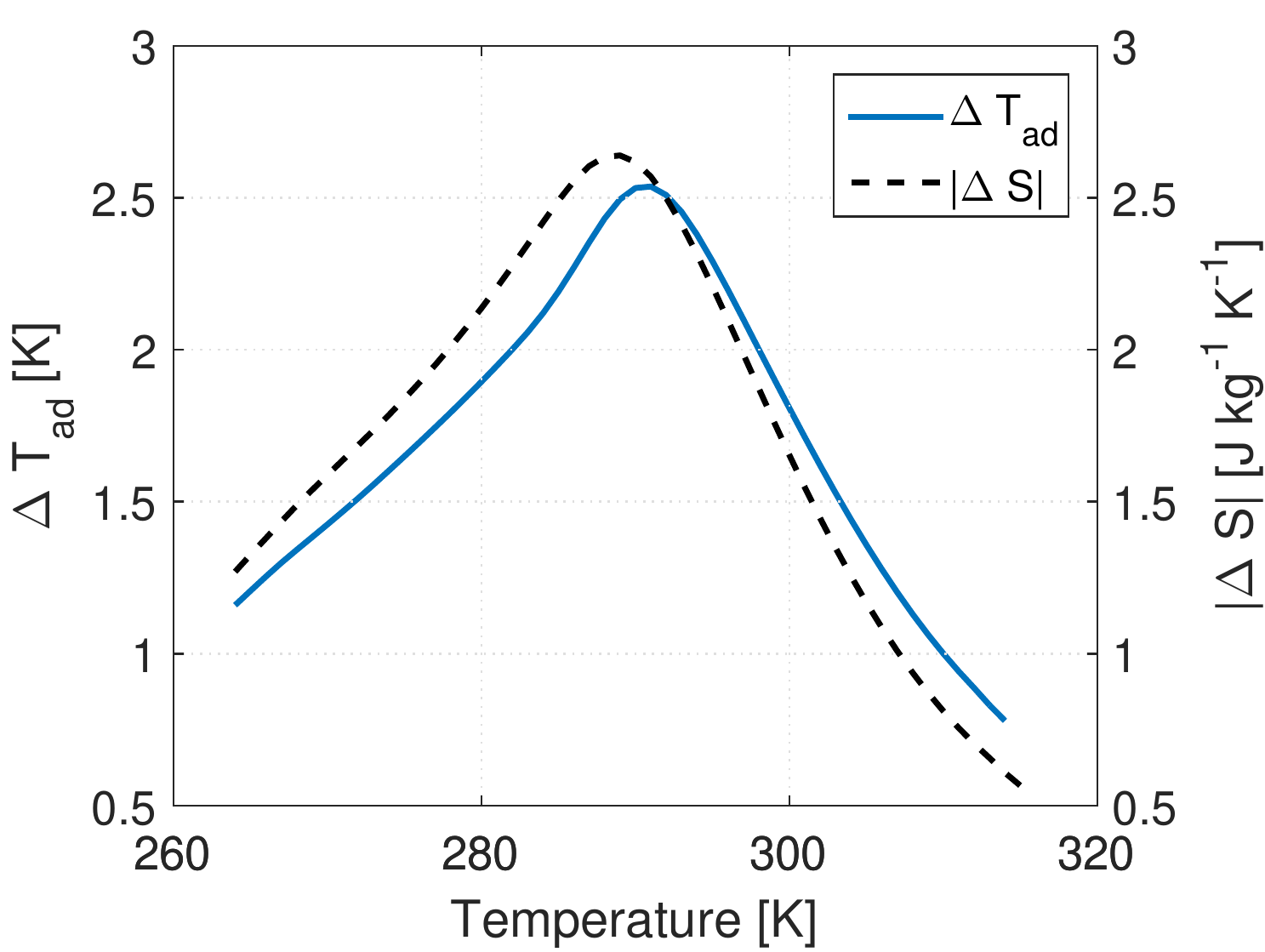}
    \caption{The adiabatic temperature change (solid line) and the isothermal entropy change (dashed line) as a function of temperature with a change in internal magnetic field from 0 to 1.0 T. The data used in the model span the temperature range given in the figure and ranges from 0 to 1.4 T internal field. Even though the calculations done in this paper assume applied fields up to 1.5 T, the internal field in the AMR is never above the limits of the dataset due to demagnetization effects in the regenerator.}
  \label{Fig_comm_Gd}
\end{figure}

We therefore consider a regenerator consisting of randomly packed spheres made of commercial grade Gd. The density of Gd is $\rho_\n{s}=7900$ kg m$^{-3}$ and the porosity of the regenerator is assumed constant at $\epsilon = 0.36$ and the sphere size is homogeneous throughout a given regenerator. Finally, edge or boundary effects such as channeling are ignored. The magnetocaloric properties are plotted in Fig. \ref{Fig_comm_Gd}. The adiabatic temperature change is notably smaller than often reported in literature \citep{Dankov_1998} as the sample was of commercial grade Gd with unknown purity.

In order to calculate the cost of a magnetic refrigeration device, the performance of the device needs to be known. Here, we use a numerical model to calculate the cooling power for a fixed temperature span of $\Delta{}T=30$ K and a hot side temperature of 300 K for a regenerator with cylindrical cross section. The calculated cooling power is a function of several parameters in the AMR model, i.e. magnetic field, $\mu_0H$, particle size, $d_\n{par}$, length of the regenerator, $L$, cross sectional area, $A_\n{c}$, frequency of operation, $f$, and thermal  utilization:
\begin{equation}
\varphi\equiv\frac{\dot{m}_0c_\mathrm{f}}{2fLA_\n{c}(1-\epsilon)\rho_\n{s}c_\n{s}}.
\end{equation}
The average mass flow rate during one of the blow periods is denoted $\dot{m}_0$ and the specific heat is $c$ with subscript f for fluid and s for the solid regenerator material. The parameters were varied as given in Table \ref{Table.AMR_parameters}, resulting in a set of 38,880 simulations.

\begin{table*}[!t]
\begin{center}
\caption{The parameters varied in the AMR model.}\label{Table.AMR_parameters}
\begin {tabular}{lrrrr}
Parameter & Minimum & Maximum & Step size & Unit\\ \hline
Magnetic field, $\mu_0H$   & 0.8 & 1.5 & 0.1  & T\\
Particle size, $d_\n{par}$ & 0.1 & 0.5 & 0.05 & mm \\
Length of regenerator, $L$ & 10  & 100 & 10   & mm \\
Cross sectional area $A_c$ & \multicolumn{3}{c}{1e-5/L}   & mm$^2$ \\
Frequency of operation     & \multicolumn{3}{c}{[1 2 4 6 8 10]} & Hz \\
Utilization                & 0.2 & 1.0 & 0.1 & -
\end {tabular}
\end{center}
\end{table*}

The numerical AMR model solves the coupled partial differential equations describing heat transfer in the fluid and solid, respectively. These are given by:
\begin{eqnarray}
c_\n{f}\rho_\n{f}\epsilon\left(\parde{T_\n{f}}{t}+u\parde{T_\n{f}}{x}\right)&=&\parde{}{x}\left(k_\n{disp}\parde{T_\n{f}}{x}\right)-ha_s\left(T_\n{f}-T_\n{s}\right)\nonumber\\
&&+\left|\frac{\Delta p\dot{m}}{\rho_\n{f}LA_\n{c}}\right|\label{eq_reg_fluid}\\
c_\n{s}\rho_\n{s}(1-\epsilon)\parde{T_\n{s}}{t}&=&\parde{}{x}\left(k_\n{stat}\parde{T_\n{s}}{x}\right)+ha_s\left(T_\n{f}-T_\n{s}\right)\nonumber\\
&&-\rho_\n{s}T_\n{s}\parde{s}{H}\parde{H}{t}.\label{eq_reg_solid}
\end{eqnarray}
The fluid flow is along the $x-$direction. The pressure drop and mass flow rate to a given time $t$ are denoted $\Delta p$ and $\dot{m}$, respectively. The specific surface area of the packed spheres is $a_s=6\frac{1-\epsilon}{d_\n{par}}$ while the convective heat transfer coefficient describing heat transfer from the surface of the spheres to the fluid is denoted $h$. The thermal conductivity in the fluid is described by an effective value including the effect of dispersion ($k_\n{disp}$) while the conductivity of the spheres is assumed equivalent to the static conduction of the regenerator ($k_\n{stat}$). The correlations for these parameters as well as for the convective heat transfer coefficient and the pressure drop as a function of mass flow rate and regenerator aspect ratio are provided alongside the numerical implementation details in \citet{Nielsen2013c}. A detailed discussion of the derivation of the active magnetic regenerator equations can be found in \cite{Engelbrecht2008}. The boundary conditions at the hot and cold end are constant temperatures ($T_\mathrm{hot}$ and $T_\mathrm{cold}$, respectively). The rest of the regenerator is assumed adiabatic with respect to the ambient, i.e. parasitic losses are neglected. The regenerator equations (\ref{eq_reg_fluid}--\ref{eq_reg_solid}) are discretized in space using a $2^\mathrm{nd}$ order finite difference approach and solved in time using the fully implicity scheme. The details are provided in \cite{Nielsen2013c}. The magnetic field profile and the flow profile are assumed trapezoidal in time with zero no-flow time. The ramp between the hot and cold blow periods is 5 \% of the total blow time.

The magnetocaloric effect is included as a source term in the equations above. At each time step the derivative of the absolute entropy with respect to magnetic field, $\parde{s}{H}$, is found as a function of temperature and local magnetic field. This magnitude of the field is found through solving the following equation iteratively in each timestep and at each spatial location:
\begin{equation}
H=H_\n{app}-NM(T,H),
\end{equation}
where $H_\n{app}$ is the applied magnetic field and $M(T,H)$ is the magnetization of the regenerator material. The average demagnetization factor, $N$, is found by combining the overall shape of the regenerator (which is cylindrical) and the approximation for a porous medium \citep{Bleaney1941}:
\begin{equation}
N=\frac{1}{3}+(1-\epsilon)(N_\n{cyl}-\frac{1}{3}).
\end{equation}
The demagnetization factor for a cylindrical shape, $N_\n{cyl}$, is a function of the diameter and length of the cylinder and may be found in \citet{Joseph1966}.
\clearpage
All simulations considered a cylindrical regenerator with a volume of 10 mL. This means that the aspect ratio of the regenerator is decreased as the length of the regenerator increases. As the effect of geometrical demagnetization is included in the AMR model, the considered system cannot be ``scaled'' to produce an arbitrary cooling load. In general scaling a system involves keeping the regenerator length constant and increasing the radius of the regenerator. This would change the demagnetization factor of the regenerator thus leading to a different performance of the AMR. Thus only regenerators with aspect ratios as given by the regenerator length in Table \ref{Table.AMR_parameters} and the volume of 10 mL are considered here. Such regenerators have an MCM mass of 47 g. In order to reach a higher cooling capacity than these 47 g can provide, one would explicitly have to build several systems, each with 47 g (i.e. a regenerator volume of 10 mL). However, in the following we do assume a smooth scaling of the cooling power for a given device. This inevitably leads to optimized devices with masses that are non-integer multiples of the base regenerator modeled. The optimization should then be continued with the found optimal regenerator volume (or mass) as the base regenerator. This must be,  however, a second order effect in terms of the resulting cost and we have therefore chosen not to take this further step in the analysis.

\section{Total cost of an AMR}
The cost of an AMR refrigeration device will be composed of the cost of the building blocks and the cost for operating the device. Both factors are included in the analysis presented here. Actual manufacturing, transportation and maintenance and auxillary systems are ignored.

The cost of the building blocks is assumed to consist of the price of the magnetocaloric material and the price of the permanent magnet material, as the remaining parts will in general be relatively cheap. In the following analysis we only consider devices where the magnetic field is supplied by a permanent magnet, as these are both the most common and the most cost-effective devices \citep{Bjoerk_2010b}. The cost of operating the device is given by the power consumed by the device multiplied by the electricity cost and by the lifetime of the device. In order to select the device with the lowest overall cost, the building cost and the operating cost must be optimized simultaneously. Before considering how to minimize the total cost, we first consider how to determine the building cost and the cost of operation for a magnetic refrigeration device.

\subsection{Operating cost}
The running cost of the magnetic refrigerator is calculated on the basis that the magnetic refrigerator is running continuously, albeit in different modes depending on the cooling load. This is in contrast to the operation of a compression-based refrigeration unit, which usually runs infrequently but provides a high temperature span and cooling power when it does run. We consider a system that will have to provide a high cooling load, $Q_\n{high}$, for a given percentage of the operating time and subsequently a lower cooling power, $Q_\n{low}$, for the remainder of the operating time, comparing this to one with a constant average load of $Q_\n{av}$.

As stated above the cost of operating the device is given by the power consumed by the device multiplied by the cost of electricity and by the lifetime of the device. The price of electricity varies a lot from country to country and depending on the pattern of usage. Here the cost of electricity is taken to be 0.1 \$ kWh$^{-1}$, relevant for, e.g., the United States \citep{Power_2015}, China and India, while many European countries have higher prices. The power needed to operate the device is given by the COP of the device once the cooling load is known. Traditionally, one would select an AMR that operates at the highest possible COP thus reducing the operating cost. However, doing this will disregard the size of the AMR and the size of the applied magnetic field. As these two factors have a significant influence on the building cost of the magnetic refrigeration device, the approach of minimizing the operation cost alone is invalid.

Instead, the operating cost has to be calculated for every single device and for every single operating condition considered. It should then be combined with the building cost in order to find the lowest overall price. However, only some of the AMR parameters will influence the building cost of a device. Of the parameters given in Table \ref{Table.AMR_parameters}, the magnetic field, particle size and length of the regenerator are inherent physical parameters of the regenerator that cannot be changed once it has been built. An AMR with a given set of values of these three parameters is here coined a ``device''. The two remaining AMR parameters, the frequency and the utilization, can be adjusted for an AMR in operation and do not affect the building cost. These are the operating parameters that will be adjusted to switch between the different cooling powers required for the device.

When adjusting the operating parameters, one can either adjust the frequency and utilization such that the produced cooling power is exactly as required. Another alternative is to adjust the frequency and utilization to a slightly higher cooling power, but with a possibly higher COP, and then compensate the too high cooling power with an electric heater. This will of course lower the total COP, by an amount as given in Eq. (\ref{Eq.COP_lower}), but it cannot be ruled out a priori that this is not a viable alternative.
\begin{equation}\label{Eq.COP_lower}
  COP_\n{With\;heater} = \frac{Q_\n{desired}}{\frac{Q_\n{c}}{COP}+Q_\n{c}-Q_\n{desired}}
\end{equation}
Here, the COP of the AMR, when the heater is added, is a function of the cooling power, $Q_\n{c}$, and COP of the device without a heater, and the desired new cooling power $Q_\n{desired}$. The COP of the device without a heater is simply given by $COP = Q_\n{c}/W$, where $W$ is the work consumed. However, for all considered AMR parameters here, using a heater results in a lower performing device than merely adjusting the frequency and utilization, so this is not a viable approach.

As mentioned, a total of 38,880 systems were modelled. This corresponds to 720 devices, i.e. an AMR with fixed magnetic field, particle size and length. The minimum operating cost for each of these 720 devices must be determined. This is done using the following scheme given for each device and each possible set of operation parameters (frequency and utilization) for that device:
\begin{itemize}
  \item Choose a device
  \item Choose a set of operating conditions (frequency and utilization)
  \subitem The mass of MCM needed for the chosen device to reach $Q_\n{high}$ is determined through linear scaling and the COP is determined.
  \subitem Determine the subset of operating conditions (frequency and utilization) where this scaled device delivers $Q_\n{low}$.
  \subitem Find the highest COPs on this subset.
  \subitem The power consumption is then found from the COPs at $Q_\n{high}$ and $Q_\n{low}$, weighted by the fraction of operation time.
  \item Repeat above for all values of frequency and utilization for the given device.
  \item Repeat above for all devices
\end{itemize}
It is noted that the subset of operating conditions will have at least one element with $f>0\; \n{Hz}$ and $\varphi>0$ since the cooling power of an AMR device is a continuous function of both these variables and that it must be zero when $f=\varphi=0$. A set, $(f_1,\varphi_1)$ that fulfills $Q(f=0,\varphi=0)=0<Q_\n{low}(f_1,\varphi_1)<Q_\n{high}$ must therefore exist.

\begin{figure}[!t]
  \centering
  \includegraphics[width=1\columnwidth]{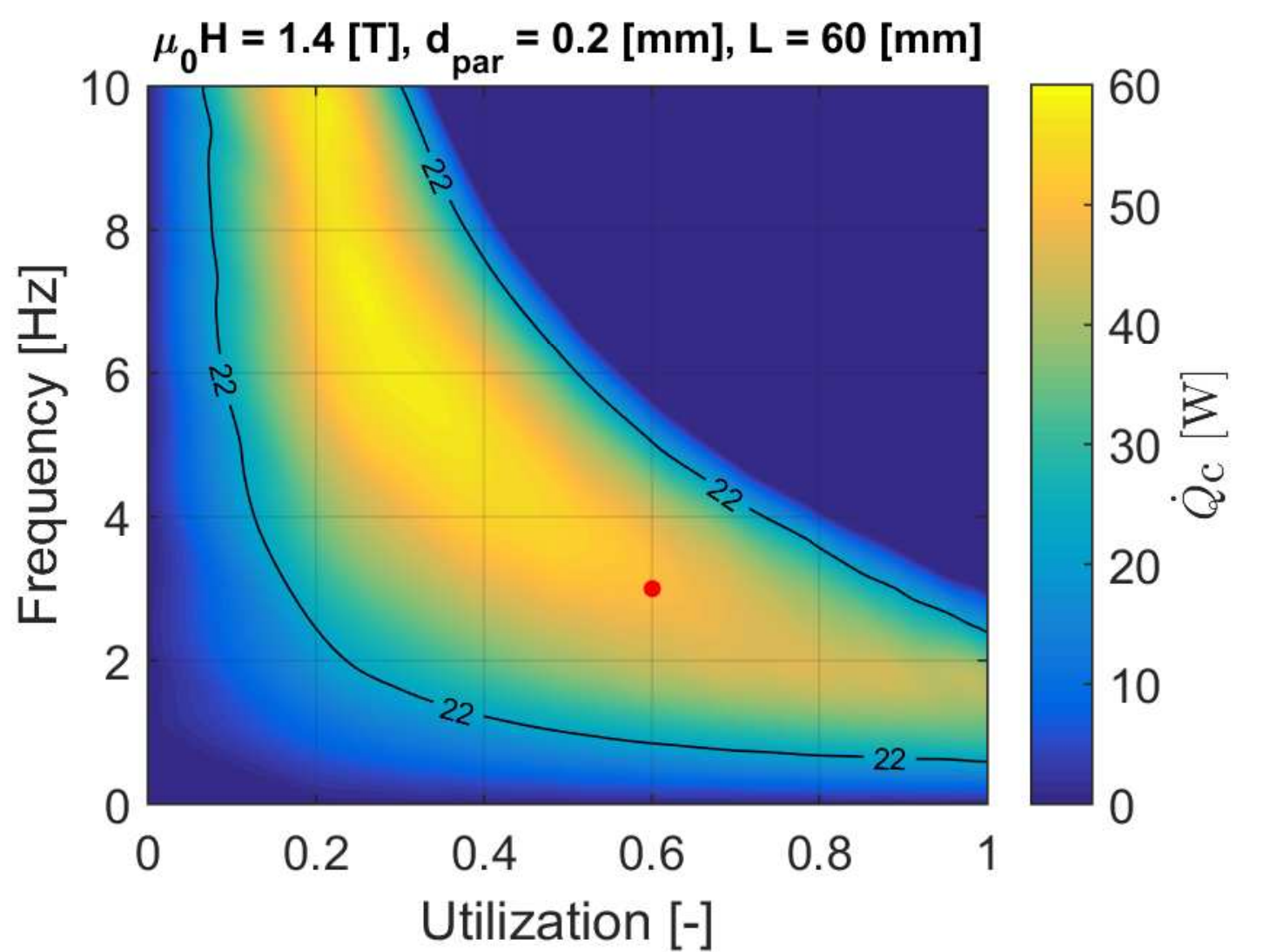}
  \caption{The cooling power as a function of utilization and frequency for an AMR with $\mu_{0}H = 1.4$ T, $d_\n{par}=0.2$ mm and $L=60$ mm. A given point, here a utilization of 0.6 and a frequency of 3 Hz, has been selected. The mass of the AMR is scaled to produce a cooling power of 50 W in this point. The contour of $Q_\n{low}$, here 22 W, is also shown. The point with lowest COP on this contour is chosen for the operating condition of $Q_\textrm{low}$.}
  \label{Fig_Example_util_freq}
\end{figure}

 Consider a cooling load of 50 W for 10 \% of the time and 22 W for the remaining 90 \% of the time. First a device with a given magnetic field, particle size and length is selected. Then the mass of the device needed to obtain $Q_\n{high}$ is determined for all values of the frequency and utilization. As an example, consider a frequency and utilization of 3 Hz and 0.6, respectively. For these values the simulated device with a volume of 10 mL produces 20 W. This means that in order for the device to be able to produce the desired 50 W, it needs to be a factor of 2.5 larger. Having determined this, the possible values of frequency and utilizations that allow this scaled device to produce $Q_\n{low}$, here 22 W, are found, as also shown in Fig. \ref{Fig_Example_util_freq}. Of these the operating condition with the highest COP is selected and this combined with the COP at the 50 W is used to calculate the operating cost, for that given frequency and utilization, here 3 Hz and 0.6, respectively. This is then repeated for all values of frequency and utilization, until the lowest operating cost and the mass of every device is known as a function of frequency and utilization.

\subsection{Building cost}
The building cost of a magnetic refrigeration unit is determined primarily by the amount of magnetocaloric material and magnet material. How to determine the amount of magnetocaloric material needed for producing the desired cooling power for a given device and operating parameters is described above. An increase in cost is expected for decreasing particle size and increasing homogeneity of the production of spherical particles. These additions to the cost are not considered further as the production of Gd spheres is very limited and thus assessing the price in a future production is associated with considerable uncertainty. Thus the remaining factor to determine is the cost of the magnet material.

As described in \cite{Bjoerk_2011d} the amount of magnet material needed to produce a given magnetic field can be calculated from the mass of the magnetocaloric material alone, by knowing the remanence and density of the permanent magnets, and the figure of merit, $M^{*}$, for the magnet system. As the latter is known to vary from 0 to 0.25, the price can ideally be determined as a function of the effectiveness of the magnet system. However, this approach does not consider that the regenerator cannot be completely enclosed in an assembly of permanent magnets, in order to reduce flux leakage, but that the ends of the regenerator must usually be left open to allow fluid flow to enter and exit the regenerator.

\begin{figure}[!t]
  \centering
  \includegraphics[width=1\columnwidth]{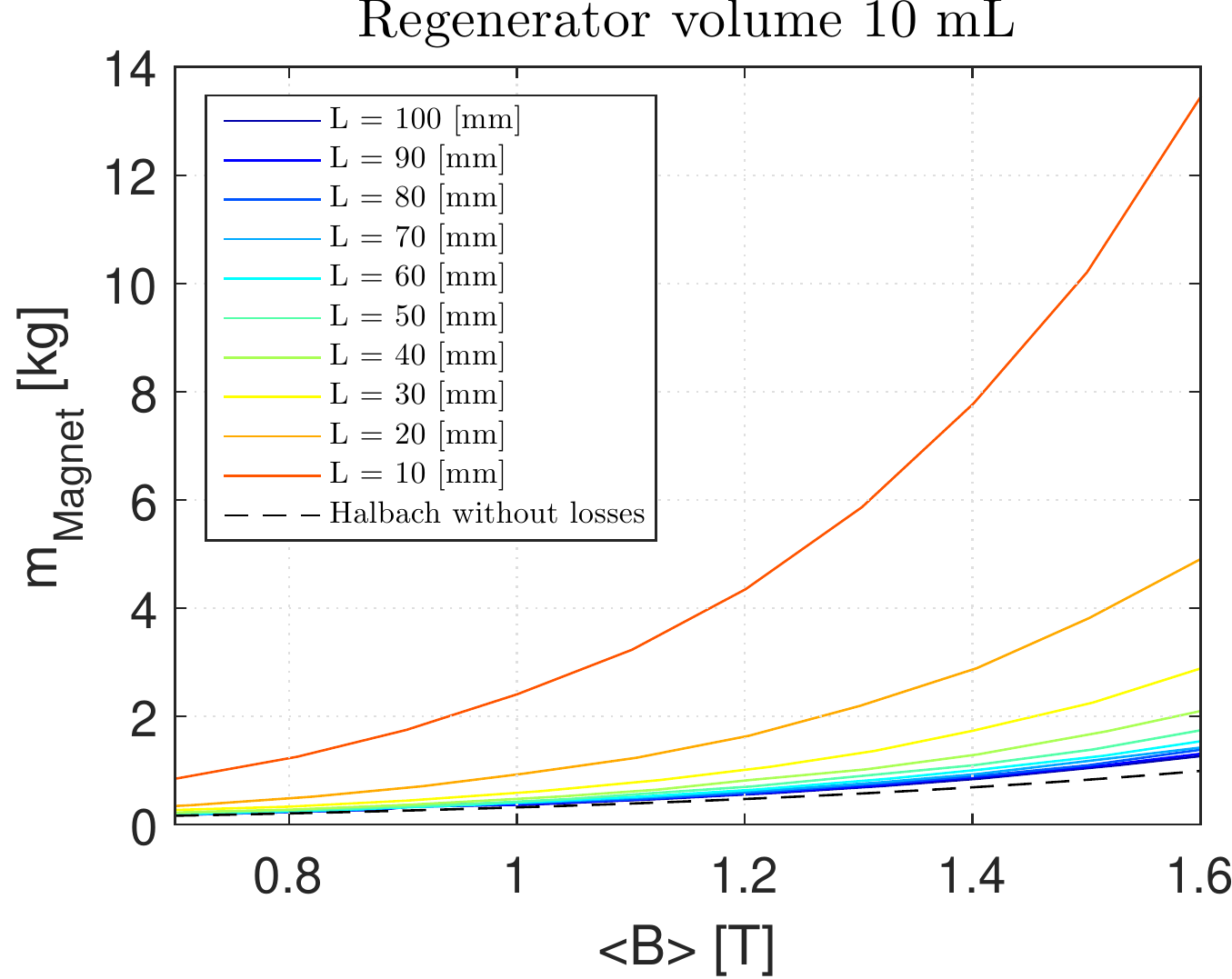}
  \caption{The mass of the magnet needed to provide a given mean magnetic field in a volume of 10 mL, with varying length of the regenerator, i.e. varying cross-sectional area. The theoretical case without any losses through the ends of the cylinder is also shown.}
  \label{Fig_Mass_magnet_Ac}
\end{figure}

Therefore, we consider instead the well known Halbach cylinder design \citep{Mallinson_1973,Halbach_1980} with a finite length as the basis for the permanent magnet system. Close to 40,000 simulations of the Halbach design were conducted in order to determine the minimum amount of permanent magnet required to generate the desired average magnetic field inside the volume of the regenerator, for the considered cross-sectional areas. The homogeneity of the magnetic field over the regenerator is not considered in this approach. We consider magnets with a remanence of 1.2 T, which is a common value for NdFeB magnets -- the most powerful type of magnet commercially available today. These have a density of $\rho_\n{mag} = 7400$ kg m$^{-3}$. The choice of a remanence of 1.2 T is to ensure that the magnets are reasonably priced, as well as to disregard possible demagnetization issues, as the coercivity of magnets decreases strongly with increasing remanence. The found magnet mass as a function of field is shown in Fig. \ref{Fig_Mass_magnet_Ac} for the different cross sectional areas considered. As can be seen from the figure, the larger the cross sectional area, the larger the losses through the ends of the magnet, and thus the more magnet material is needed to create the desired field. This will subsequently be weighted against the increase in pressure drop, and thus pumping power, for the longer regenerators.

\section{Minimizing the life-time cost of a magnetocaloric refrigerator}
The total cost of a magnetic refrigerator is given as the sum of the cost of the magnet material, the magnetocaloric material and the operating cost. This can also be put in terms of the price of magnet material, $\$/\n{kg}_\n{Magnet}$, the price of magnetocaloric material, $\$/\n{kg}_\n{Magnetocaloric}$, the price of electricity, \$ kWh$^{-1}$, and the mass of the magnet, $m_\n{Magnet}$, the magnetocaloric material, $m_\n{MCM}$, the power used by the device, $P$ and finally the lifetime of the device, $t_\n{Life}$, i.e.
\begin{eqnarray}
\n{Cost}_\n{Total} &=& \n{Cost}_\n{Magnet} + \n{Cost}_\n{MCM} + \n{Cost}_\n{Operation} \nonumber\\
                   &=& \$/\n{kg}_\n{Magnet}m_\n{Magnet} + \$/\n{kg}_\n{MCM}m_\n{MCM} \nonumber\\
                   &&+ \$/\n{kWh}\;P\;t_\n{Life}
\end{eqnarray}

These are the major factors contributing to the cost of the magnetic refrigerator. The cost of various standard components such as a motor as well as various materials for construction are assumed to be small compared to the factors mentioned above. Furthermore, the price of these standard components does not change the optimization, as their price remains constant. We consider the total cost of the refrigerator over its lifetime as a function of the price of magnet and magnetocaloric material. These values influence all factors in the equation, as e.g. a more expensive magnet price might lead to a slightly smaller regenerator but with an increased operating cost. We do not consider that any of the cost from building the AMR can be recovered, i.e. the price of the magnet and the MCM are assumed to be lost once the device has been built. This is a very conservative assumption representing a worst case scenario. In reality various recycling schemes will be able to recover at least some parts of the magnet and MCM costs \citep{Habib_2014}.

\begin{figure*}[!t]
\centering
\subfigure[24.8 W]{\includegraphics[width=1\columnwidth]{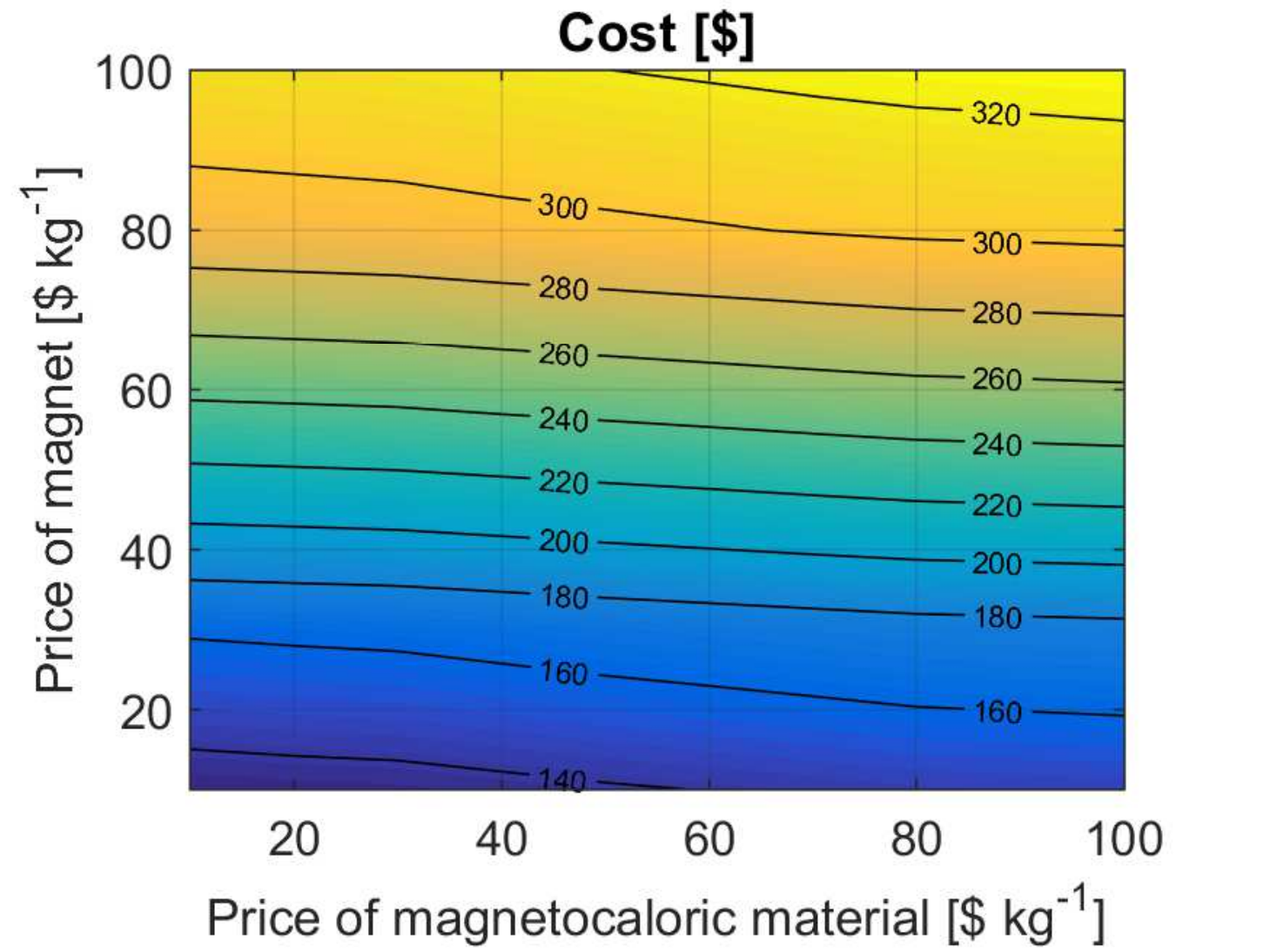}}\hspace{0.2cm}
\subfigure[50 W - 22 W]{\includegraphics[width=1\columnwidth]{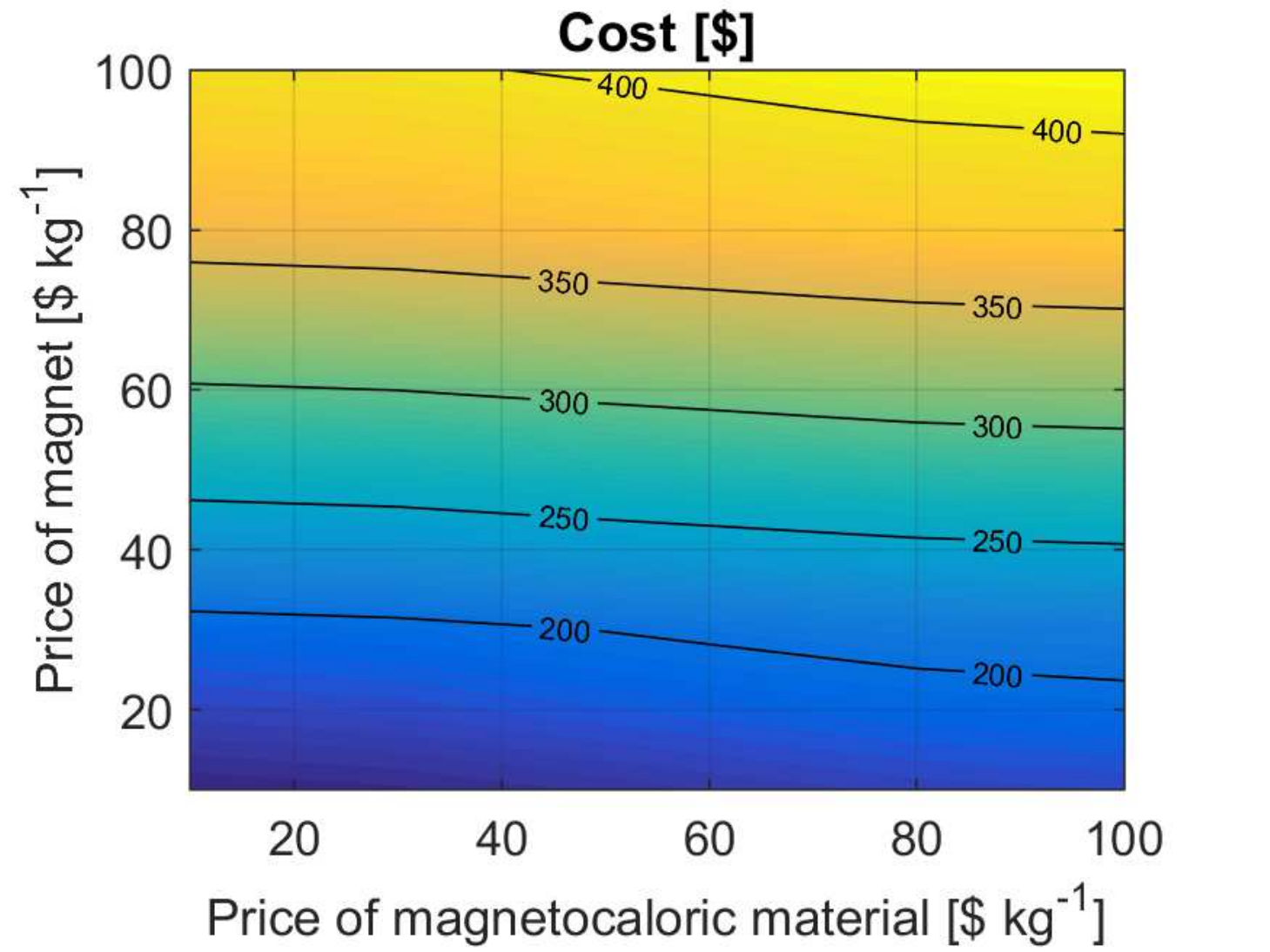}}
\caption{The total cost as a function of the price of the magnet material and the magnetocaloric material for (a) a 24.8 W refrigerator and (b) a 50 W - 22 W refrigerator}\label{Fig.Cost_total}
\end{figure*}

The total minimum cost of the magnetic refrigerator with an expected lifetime of 15 years, as a function of the price of the magnet material and the magnetocaloric material is shown in Fig. \ref{Fig.Cost_total} for the two systems described above. An expected lifetime of 15 years was chosen in accordance with \citet{Energy_2011}. The total cost over the entire lifetime of the device is seen to range from \$150 to \$400, depending on the price of the magnet material and the magnetocaloric material. For all material prices, the refrigerator running at constant load is seen to be cheaper by 15-25\%, a value increasing as a function of the price of the magnet material.

Comparing the cost of two technologies should always be done with some care, especially if the two are not at the same stage of development. But if we were to consider the lifetime cost of the vapor compression appliance in a similar approach, we could make a very rough comparison. An A$^{+++}$ compression based unit will use \$113 of power during 15 years at 8.6 W and 0.1 \$ kWh$^{-1}$. The base price of the compressor varies, but assuming a reasonable cost of about \$30 (see, e.g. \citet{Vincent_2006}) this makes the total cost of the AMR based refrigeration unit only slightly more expensive than the compressor based one. Assuming that the cost of the magnet AMR system can be recuperated at end of life, the AMR device will actually end up being cheaper than the compressor in this rough comparison.

\begin{figure*}[!t]
\centering
\subfigure[24.8 W]{\includegraphics[width=0.9\columnwidth]{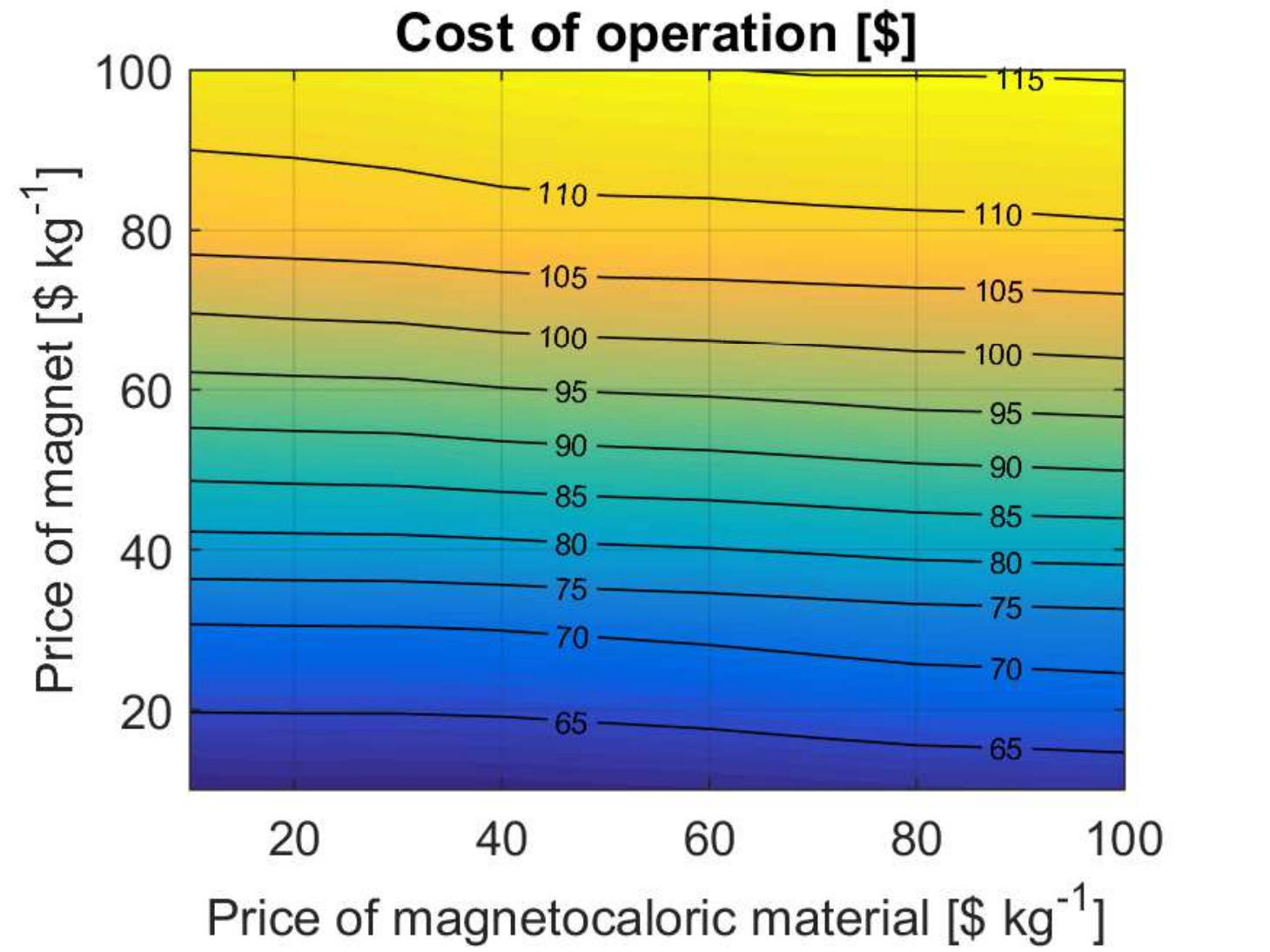}}\hspace{0.2cm}
\subfigure[50 W - 22 W]{\includegraphics[width=0.9\columnwidth]{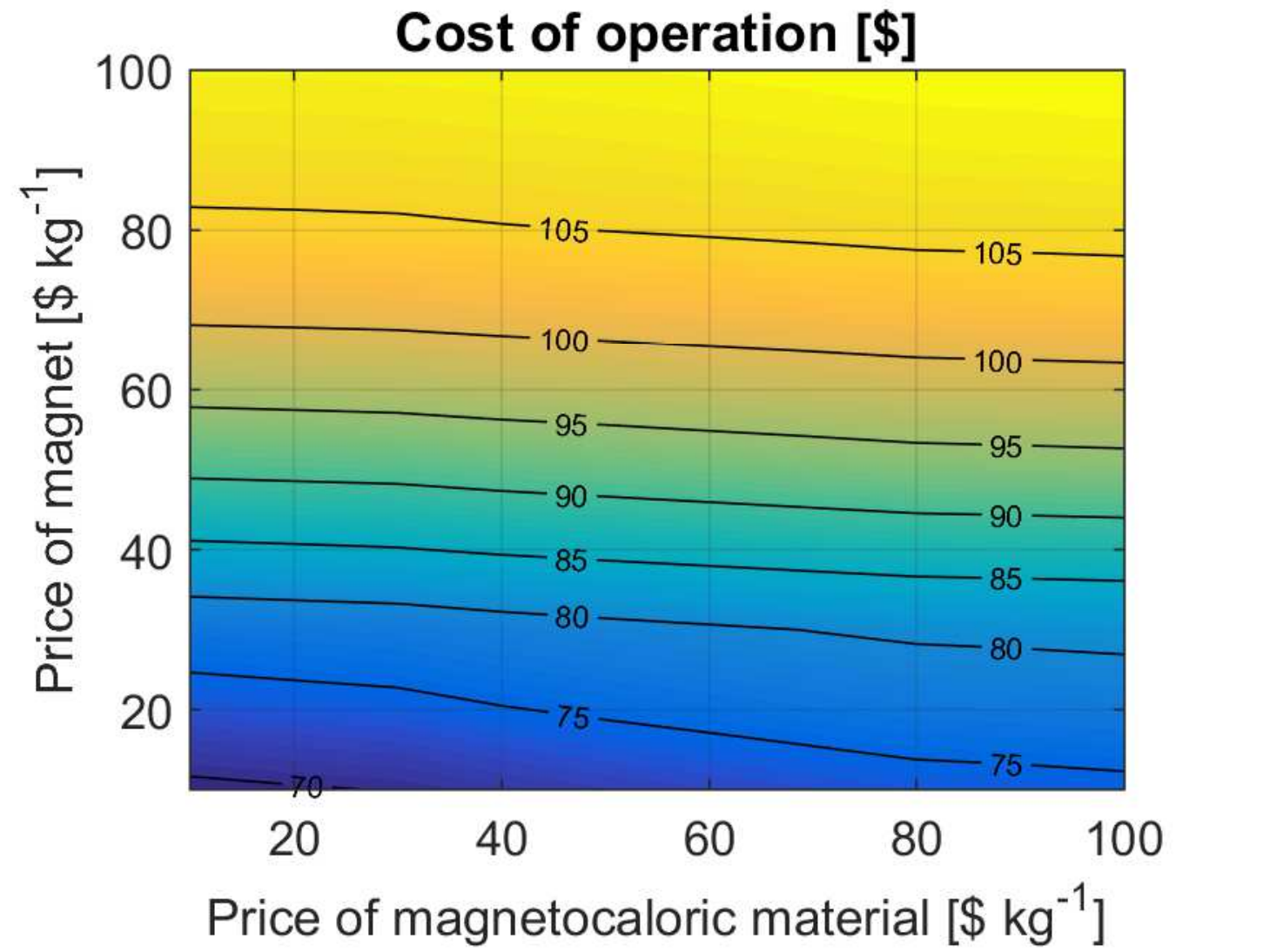}}
\caption{The cost of operating the refrigerator as a function of the price of the magnet material and the magnetocaloric material for (a) a 24.8 W refrigerator and (b) a 50 W - 22 W refrigerator}\label{Fig.Cost_ope}
\end{figure*}

\begin{figure*}[!t]
\centering
\subfigure[24.8 W]{\includegraphics[width=0.9\columnwidth]{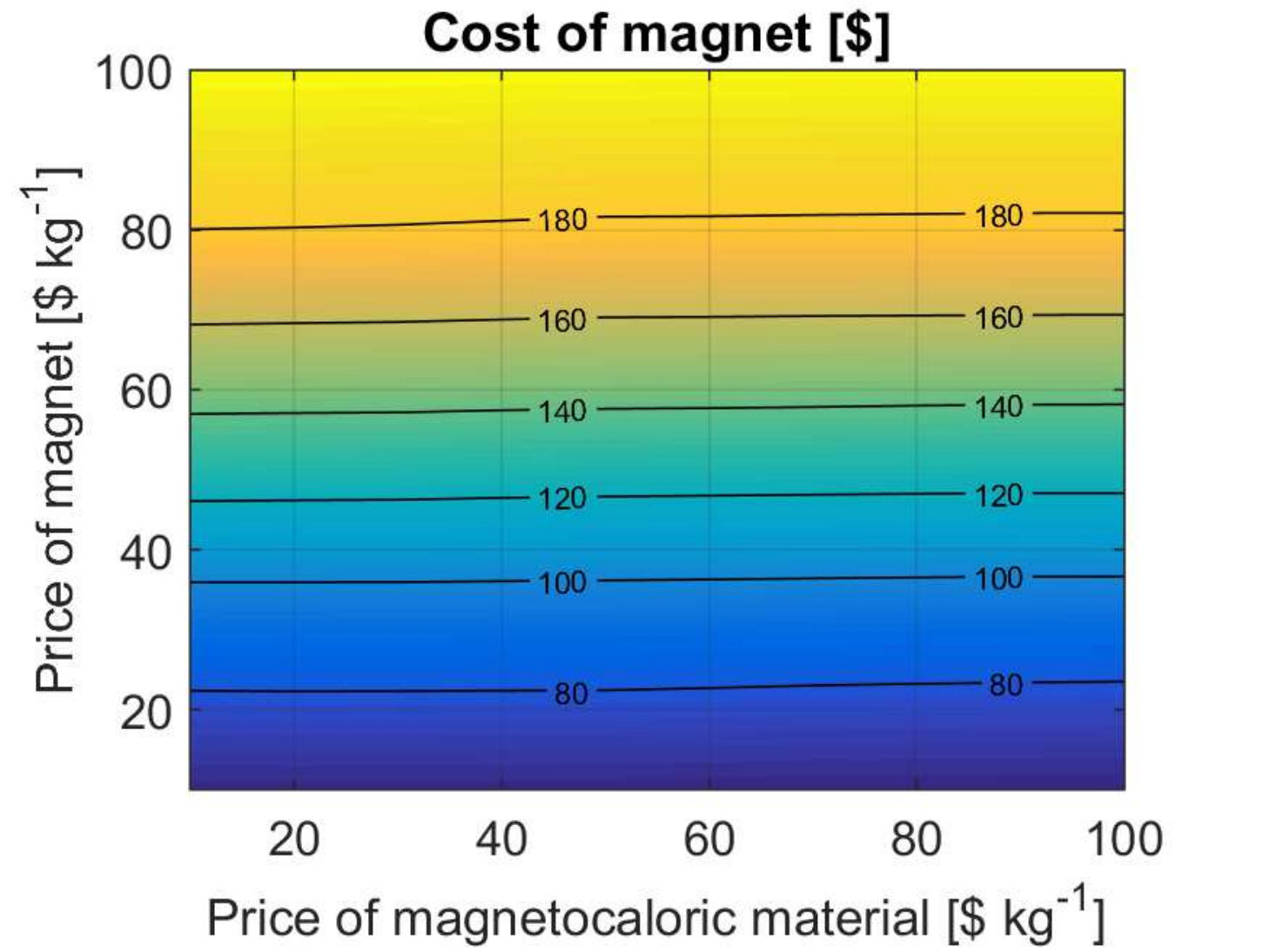}}\hspace{0.2cm}
\subfigure[50 W - 22 W]{\includegraphics[width=0.9\columnwidth]{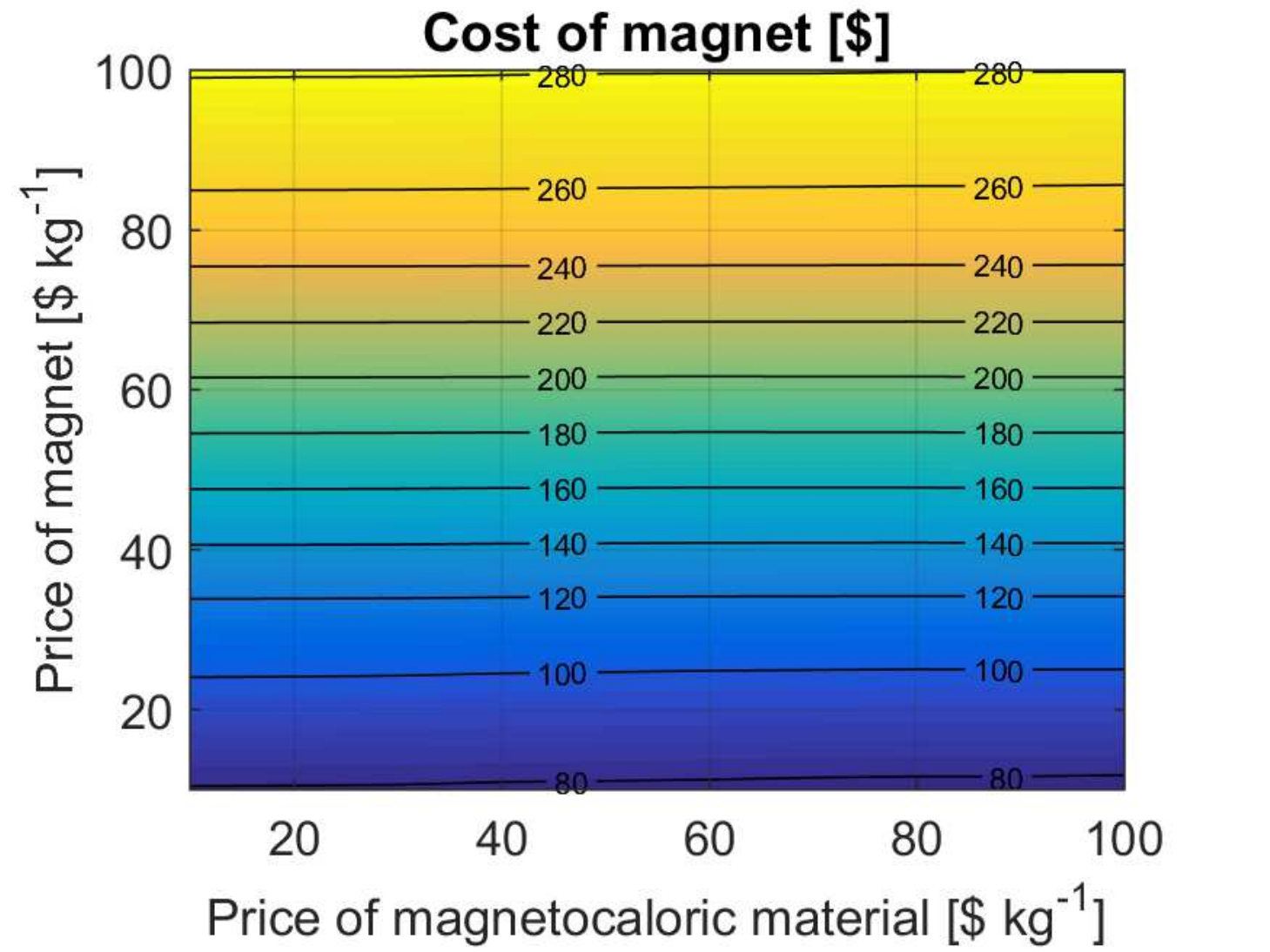}}
\caption{The cost of the magnet as a function of the price of the magnet material and the magnetocaloric material for (a) a 24.8 W refrigerator and (b) a 50 W - 22 W refrigerator}\label{Fig.Cost_mag}
\end{figure*}

\begin{figure*}[!t]
\centering
\subfigure[24.8 W]{\includegraphics[width=0.9\columnwidth]{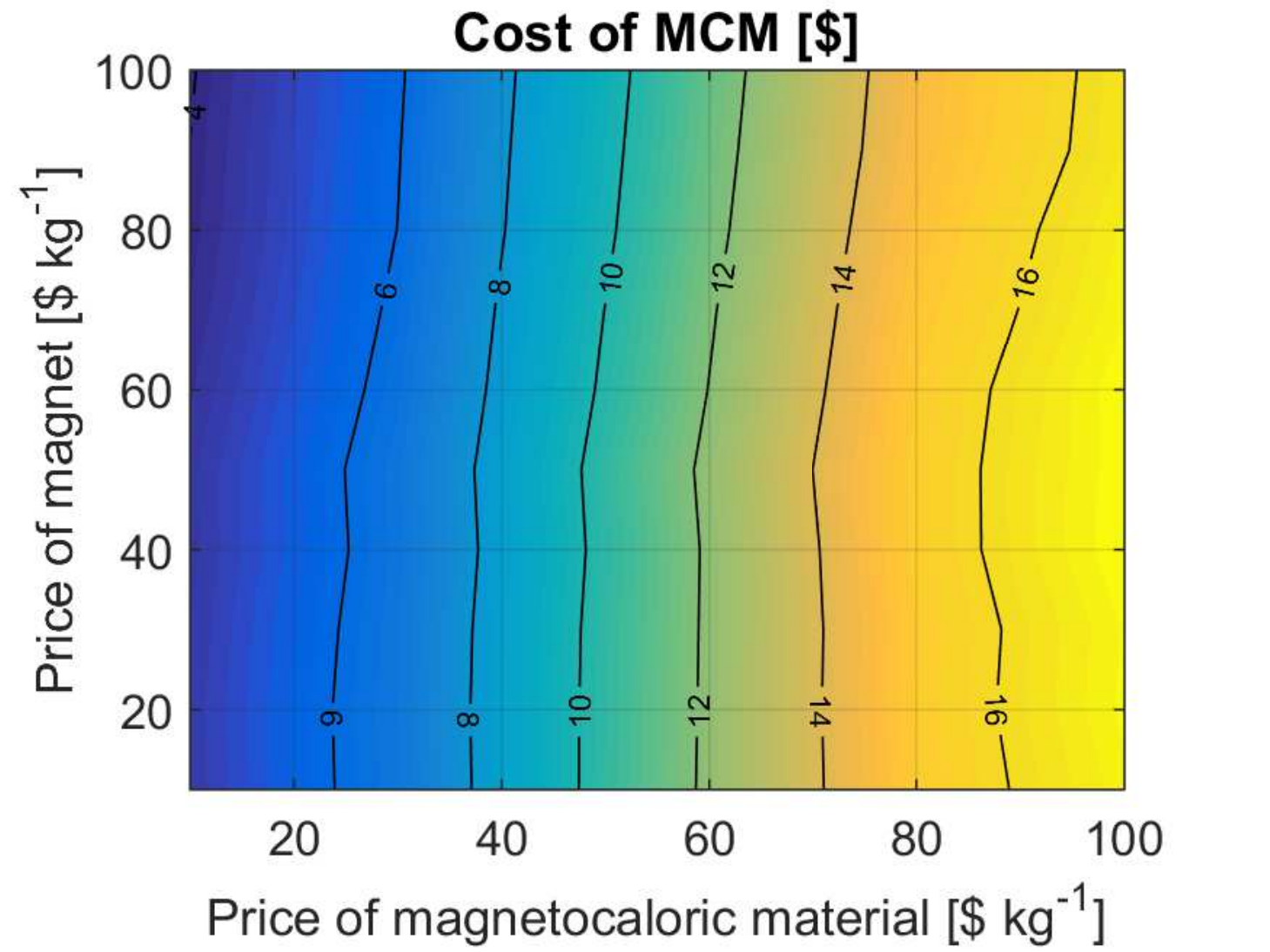}}\hspace{0.2cm}
\subfigure[50 W - 22 W]{\includegraphics[width=0.9\columnwidth]{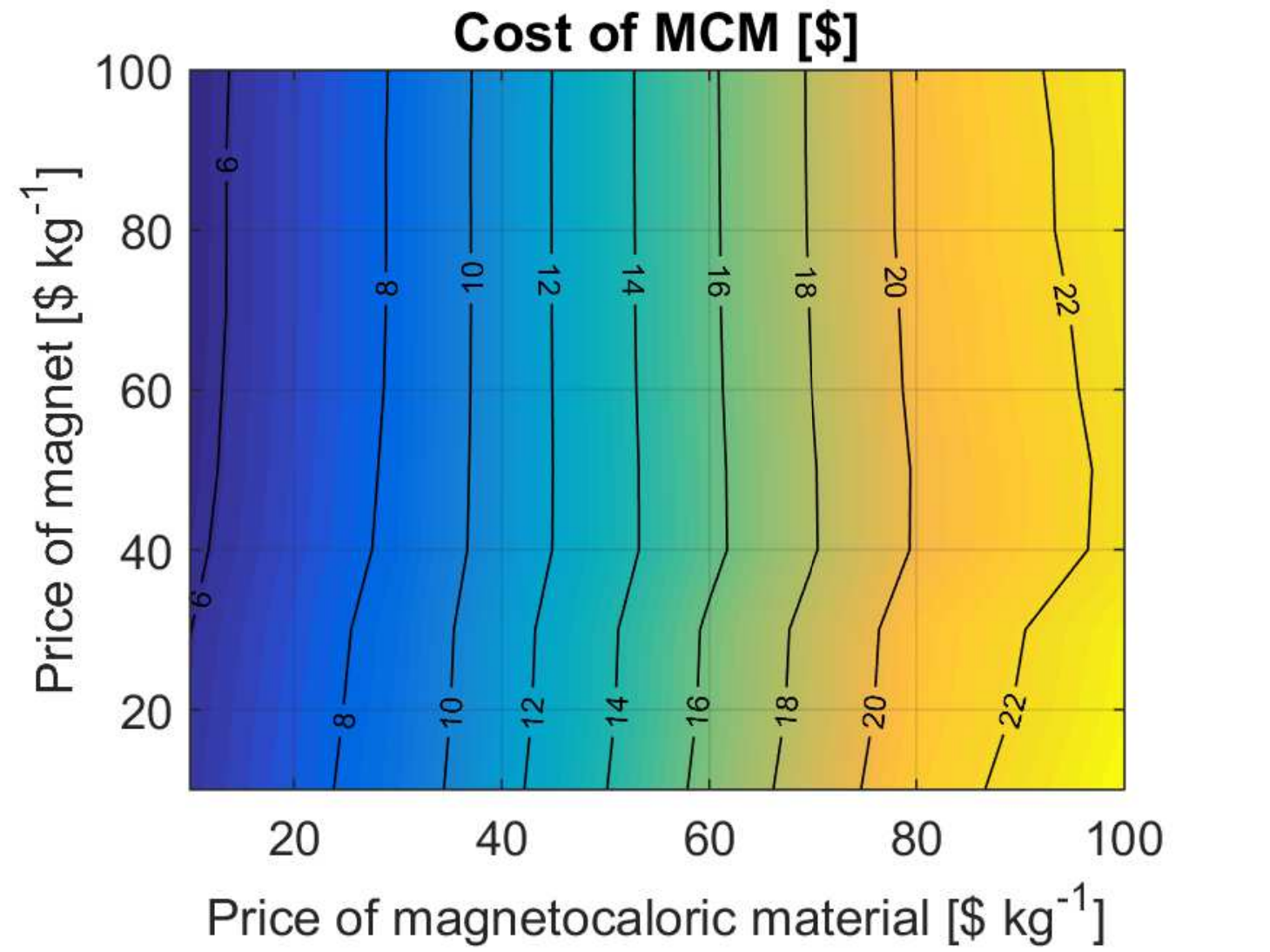}}
\caption{The cost of the regenerator as a function of the price of the magnet material and the magnetocaloric material for (a) a 24.8 W refrigerator and (b) a 50 W - 22 W refrigerator}\label{Fig.Cost_mat}
\end{figure*}

As mentioned above the total cost is the sum of the cost of the magnet material, the magnetocaloric material and the operating cost. These individual components of the total cost are illustrated in Figs. \ref{Fig.Cost_ope}, \ref{Fig.Cost_mag} and \ref{Fig.Cost_mat}, respectively, for the two systems. For both systems, the magnet is seen to be the largest factor in determining the total cost, followed closely by the cost of operation. The cost of the magnetocaloric material is seen to be almost negligible for both types of system. Interestingly, the cost of operation is in some cases lower for the 50 W - 22 W system than for the 24.8 W system. This is because a much larger magnet, with a slightly higher magnetic field and with room for a larger regenerator, is preferred for this system. This is prioritized in order to reach the specified 50 W. This also makes the magnet much more expensive for the 50 W - 22 W, which results in the increase in total cost seen for this system.

\subsection{Operating parameters}
The operating parameters and other device specific parameters, such as the mass of the magnet and the mass of the magnetocaloric material are shown in \ref{Sec_app} for the two systems, respectively. From these figures it is seen that a larger magnet is prioritized for the 50 W - 22 W system. The magnetic field is seen to be slightly larger, $\sim{}0.05$ T, the regenerator is longer, $\sim{}5$ mm, and the system is scaled so that the mass of magnetocaloric material is larger by about 0.08 kg. The particle size remains the same for the two systems, 0.225-0.245 mm. The COP for the 24.8 W system is very close to the average COP of the 50 W - 22 W system. Interestingly, a very high COP is prioritized for the 50 W - 22 W system operating at the lower cooling power, while the system has a low COP at the high cooling power. This is due to the 90 \% - 10 \% operating times of the system. The utilizations of the two systems are seen to be similar at 0.2-0.23. The frequency is, however, seen to be quite different between the two systems. The 50 W - 22 W system operates at a frequency of 7-9 Hz at high cooling load, 50 W, and at 2.5-3 Hz at low cooling load, 22 W. This is in contrast to the 24.8 W system which operates at 4.5-6 Hz continuously. Again, this is caused by the fact that the 50 W - 22 W system needs to be able to provide 50 W, and then adjust the frequency of the machine to reduce the cooling power to the 22 W cooling load operation.

\subsection{Cost as a function of expected lifetime}
\begin{figure*}[!t]
\centering
\subfigure[]{\includegraphics[width=1\columnwidth]{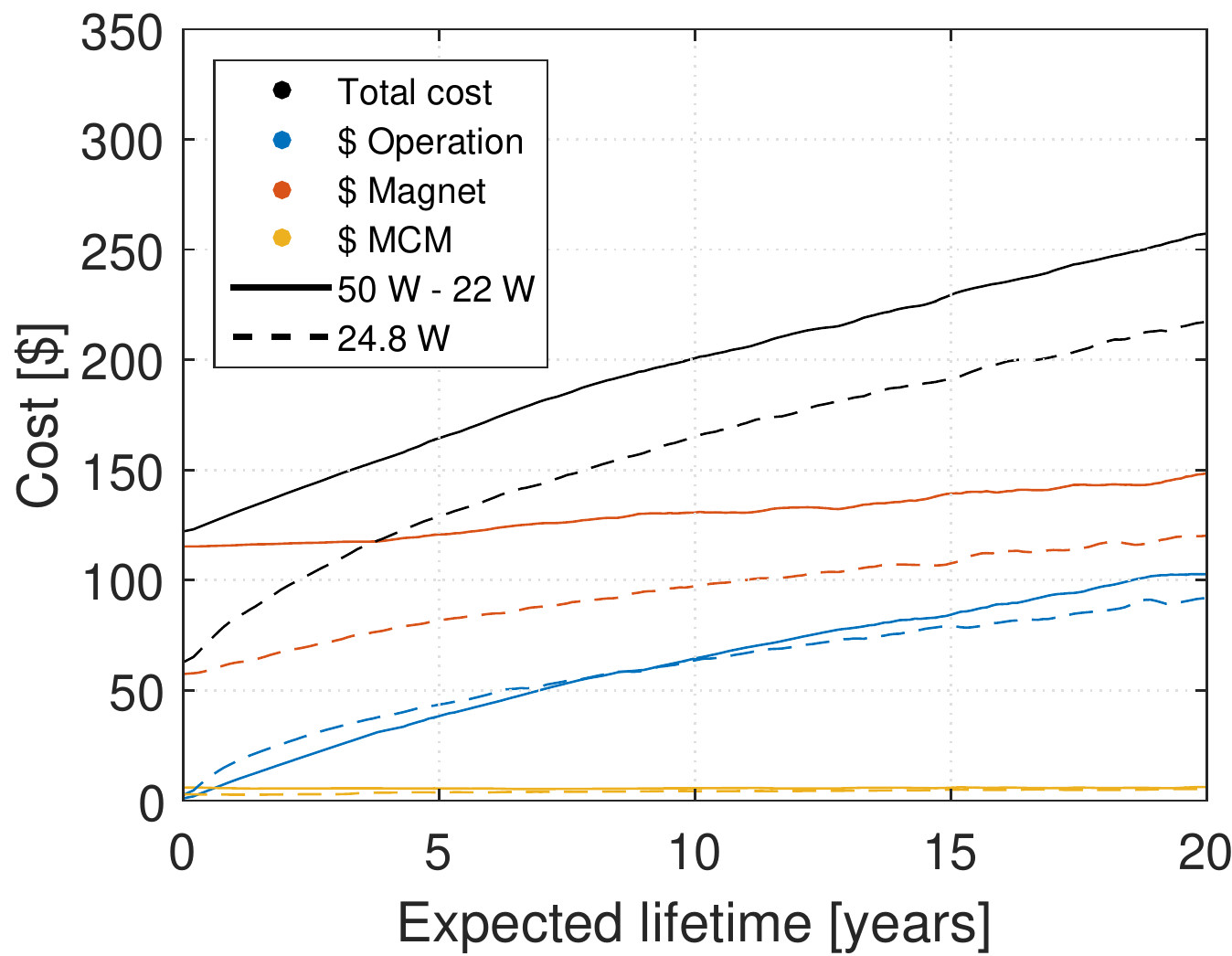}}\hspace{0.2cm}
\subfigure[]{\includegraphics[width=1\columnwidth]{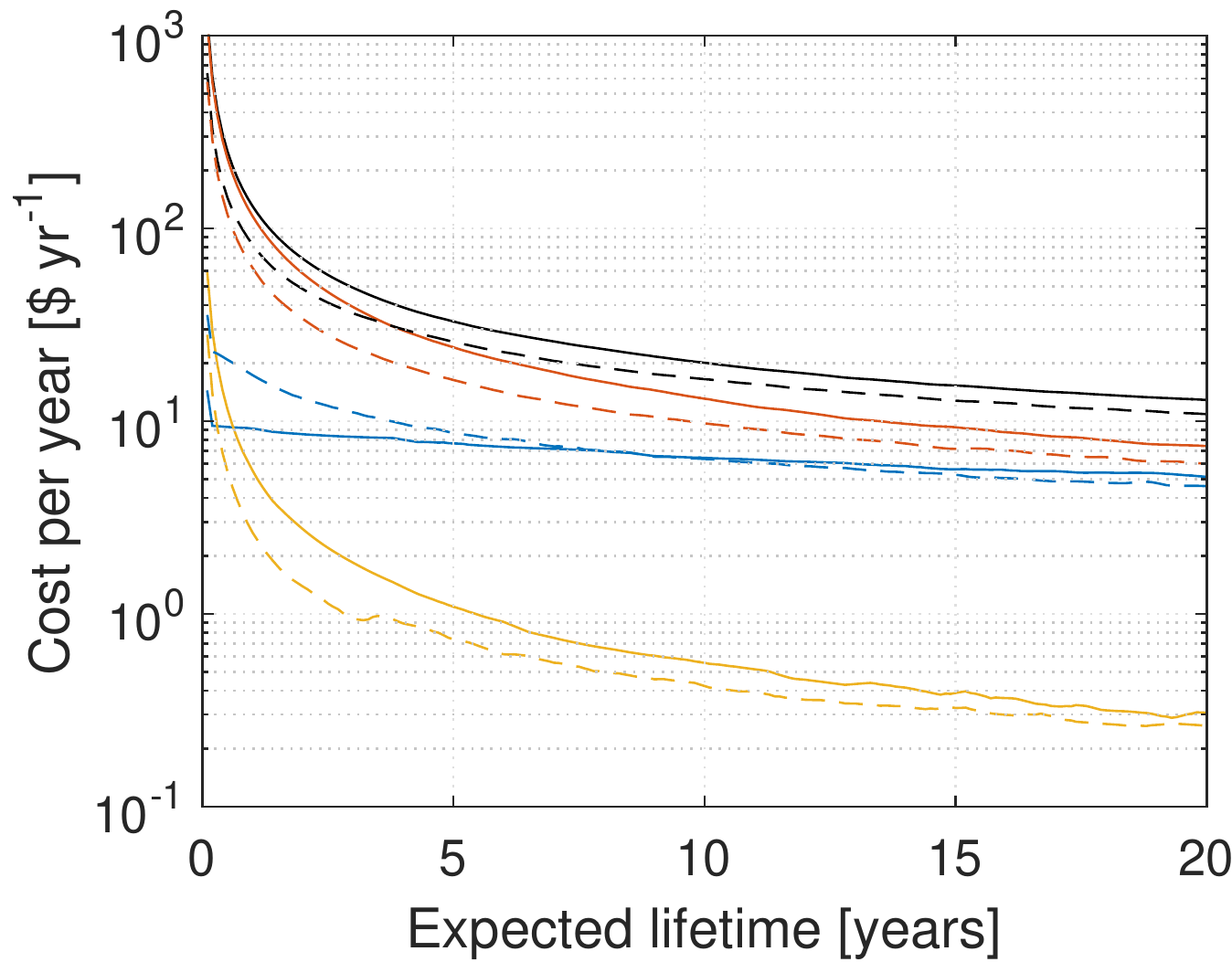}}
\caption{The (a) total cost and (b) the cost per year as a function of the expected lifetime of the AMR device for the different components. The price of the magnet material is taken to be \$40 per kg and the price of the MCM to be \$20 per kg.}\label{Fig_Cost_time_specific}
\end{figure*}

The total lifetime cost, the cost per year, and the different components of the cost, can also be examined as a function of the expected lifetime of the device. This is shown in Fig. \ref{Fig_Cost_time_specific} for the case of a price of the magnet material of \$40 per kg and the price of the MCM of \$20 per kg. As can be seen from the figure, the total cost increases with the years in operation of the device. This is expected, as the cost of operation continues to increase as the device is operating. Interestingly, the price of the magnet is also seen to increase as a function of the years of operation. This is because the longer the device is in operation, the better it is to invest in a larger magnet with a larger bore, allowing a larger cooling power, which in turn lowers the cost of operating the device. The cost of the MCM remains an insignificant contributor to the total cost. If examined as the cost per year, all costs decrease substantially as a function of time, and the cost of operation approaches a constant value.

\subsection{The optimal design and operation parameters}
The operational and device parameters for the two systems, i.e. the particle size, magnetic field, length of the regenerator, frequency and utilization must also be considered as a function of the lifetime. As shown in \ref{Sec_app}, most of the operation parameters for the magnetic refrigerator vary little as a function of the price of the magnet material and the magnetocaloric material. This turns out also to be the case as a function of the expected lifetime of the device. This is the case for the value of the magnetic field, utilization, particle size and length of the regenerator. The mean values for these parameters for the lowest cost device, for expected lifetimes from 0 to 20 years and magnet material and magnetocaloric material prices of 10-100 \$ kg$^{-1}$ is given in Table \ref{Table.Parameters_all}. The remaining parameters, i.e. the amount of magnet material and magnetocaloric material, as well as the COP, are functions of the material cost and the expected lifetime, and their optimal values must be determined based on these parameters.

\begin{table*}[!t]
\begin{center}
\caption{The optimal values for the AMR parameters. The standard deviation is given by the price of the magnet material and the price of the magnetocaloric material, both ranging from \$10 to \$100 per kg, and the expected lifetime of the device from 0 to 20 years.}\label{Table.Parameters_all}
\begin {tabular}{lrrr}
Parameter                   & 50 W - 22 W device & 24.8 W device & Unit\\ \hline
Magnetic field, $\mu_0H$    & 1.43 $\pm$ 0.05 & 1.41 $\pm$ 0.05 & T\\
Particle size, $d_\n{par}$  & 0.23 $\pm$ 0.01 & 0.23 $\pm$ 0.01 & mm \\
Length of regenerator, $L$  & 48 $\pm$ 2  & 44 $\pm$ 4   & mm \\
Utilization at $Q_\n{high}$ & 0.23 $\pm$ 0.02 & \multirow{2}{*}{0.21 $\pm$ 0.01}  & - \\
Utilization at $Q_\n{low}$  & 0.200 $\pm$ 0.001 &  & -
\end {tabular}
\end{center}
\end{table*}

\begin{figure}[!b]
  \centering
  \includegraphics[width=1\columnwidth]{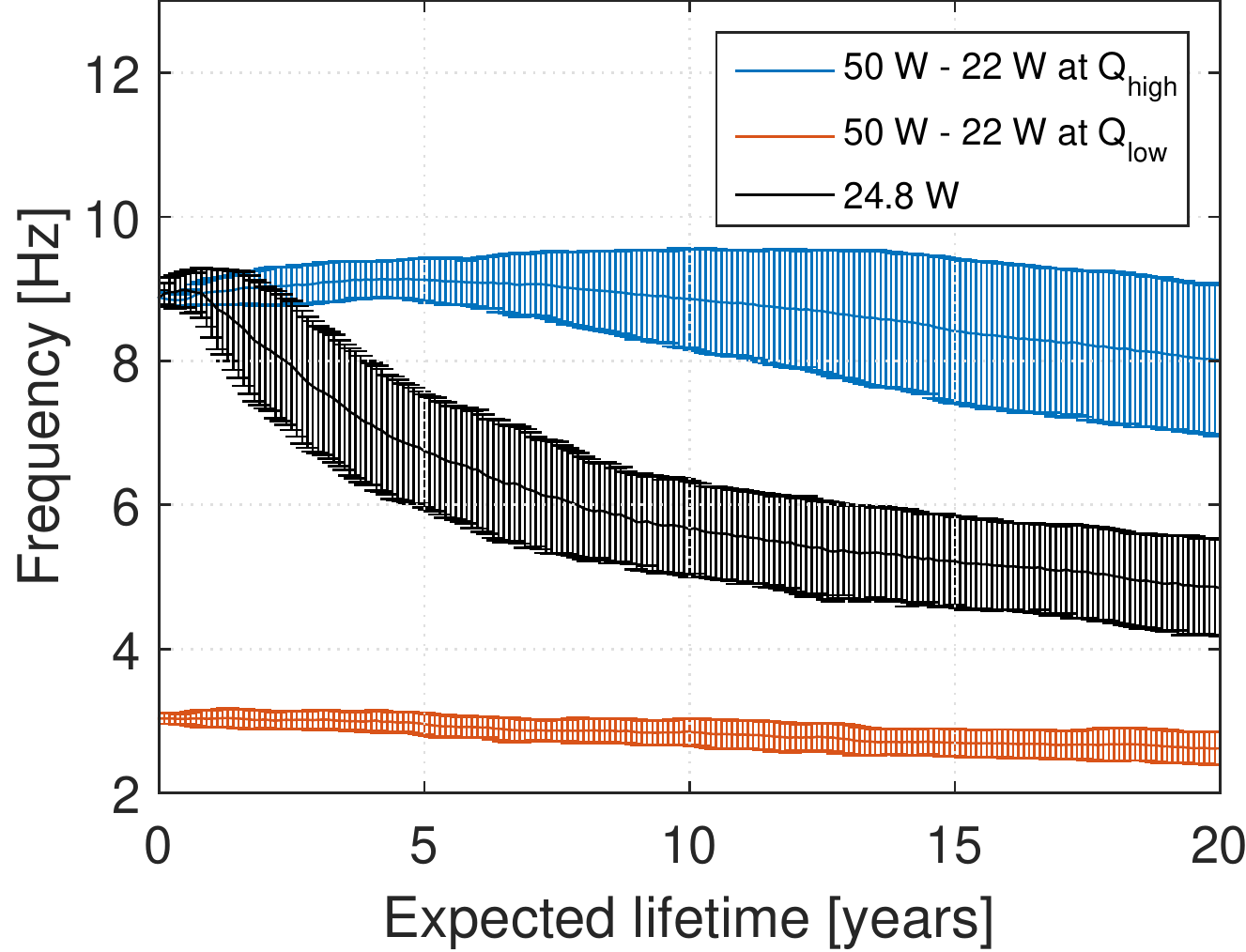}
    \caption{The frequency as a function of the function of the expected lifetime of the AMR device. The error bars are given by the value of frequency as a function of the price of the magnet material and the price of the magnetocaloric material, both ranging from \$10 to \$100 per kg.}
  \label{Fig_Cost_time_frequency_all}
\end{figure}

The operating frequency of the optimal device is a strong function of the expected lifetime of the device, and only a weak function of the price of either the magnet material or the magnetocaloric material. This is shown in Fig. \ref{Fig_Cost_time_frequency_all}, which gives the operating frequency as a function of the expected lifetime, for prices of the magnet material and the price of the magnetocaloric material ranging from \$10 to \$100 per kg. As can be seen from the error bars, the frequency is only a weak function of the price of either material. It can also be seen that for the 50 W - 22 W device, a high frequency is prioritized for the high cooling load, and a low frequency for the low cooling load. Thus the device regulates the cooling load by adjusting the frequency at which it is operating. The desired operating frequency is also seen to decrease as the expected lifetime of the device increases.

\section{Comparison with previous results and discussion}
As discussed in the introduction two previous studies of the building costs of magnetic refrigerators have been reported \citep{Bjoerk_2011d, Tura_2014}. There is a very good agreement between the values found by \citet{Tura_2014} and those reported in this work. The utilization, frequency and mass of the magnet are all in agreement. However, the COP and magnetic field found by \citet{Tura_2014} are lower than those reported in this work. The lower value of the COP is due to the larger temperature span, while the lower value of the magnetic field is due to the ``poorer'' commercial grade Gd used in this study. If a better MCM can be found and applied an increase in the performance of the AMR would be expected. The discrepancy between the values reported by \citet{Bjoerk_2011d} and those reported here is due to three reasons, each of which can be evaluated individually. The first is the use in \citet{Bjoerk_2011d} of an infinite Halbach cylinder, as compared to the optimal cylinder of finite length used in this study. From Fig. \ref{Fig_Mass_magnet_Ac} this is seen to increase the mass of the magnet by a factor between 1.5 and 2 for the optimal length of the regenerator of around 50 mm. The second effect is the inclusion of demagnetization in the AMR model. For the optimal geometry considered here, the demagnetization factor is 0.44. Finally, the last factor is that of the material data. In the previous study mean field Gd data were used. The material data used in the present study are experimentally measured properties of commercial grade Gd. The difference between the mean field Gd data and that of the data used in the present study is 1.3 K in a 1 T magnetic field.

It is also of interest to compare the AMR parameters found in this study to those reported for actual operating AMR prototype devices. Thus, we can directly compare the case of only a single $Q_\n{load}$ with previous results published in literature. However, as all of these devices are prototype devices uniquely designed and constructed, the cost of them will be very high and any comparison to the numbers in this study, optimized for mass production, will be meaningless. In \citet{Tusek_2013b} for an optimal COP configuration, the optimal AMR, disregarding magnet, is reported to have a length of 40 mm and 20 mm for 0.5 Hz and 3 Hz, respectively, and a particle size of 0.17 mm in both cases. These values are in good agreement with the values found here, even though the magnet is not considered at all in the study by \citet{Tusek_2013b}. The study by \citet{Tusek_2013b} uses mean field Gd for the MCM properties, which explains the difference in reported frequencies compared to those found here.

\section{Conclusion}
The total cost of a magnetic refrigerator was calculated, using a numerical model. The magnetocaloric material was assumed to be commercial grade Gd. Using a set of 38,880 simulations, the cost of operating and the cost of building a magnetic refrigeration unit capable of operating at 50W for 10\% of the time and 22W at 90\% of the time was determined. This was compared with a device running 24.8 W continuously. Based on these the lowest combined cost of the device was determined, as a function of the price of magnetocaloric material and magnet material. The total cost, with a device lifetime of 15 years, was found to be in the range \$150-\$400, with the cost being lowest for the 24.8 W operating device. The cost of the magnet is the dominant cost factor, followed closely by the cost of operation, while the  cost of the magnetocaloric material is almost negligible.

The optimal device and operating parameters of the magnetic refrigeration device were also determined. The optimal magnetic field was about 1.4 T, the particle size was 0.23 mm, the length of the regenerator was 40-50 mm and the utilization was about 0.2, for all device lifetimes and all considered prices of the magnetocaloric and magnet materials. The operating frequency was found to vary as a function of device lifetime. For the 50 W - 22 W system the frequency changed from 7-9 Hz at high cooling load to 2.5-3 Hz at low cooling load, in contrast to the 24.8 W system which operated at 4.5-6 Hz continuously, for a device with a lifetime of 15 years.

In a rough life time cost comparison between the AMR device and a conventional A$^{+++}$ refrigeration unit we find similar costs, the AMR being slightly cheaper, assuming the cost of the magnet can be recuperated at end of life.

\section*{Acknowledgments}
This work was in part financed by the ENOVHEAT project which is funded by Innovation Fund Denmark (contract no 12-132673). The authors acknowledge Mr. Kai Nitschmann and Dr. Carsten Wei{\ss} from BSH Hausger\"ate GmbH for valuable discussions.

\appendix
\onecolumn
\section{Operation parameters for a 24.8 W and a 50 W - 22 W system}\label{Sec_app}
\begin{figure*}[!b]
\centering
\subfigure[24.8 W]{\includegraphics[width=0.49\columnwidth]{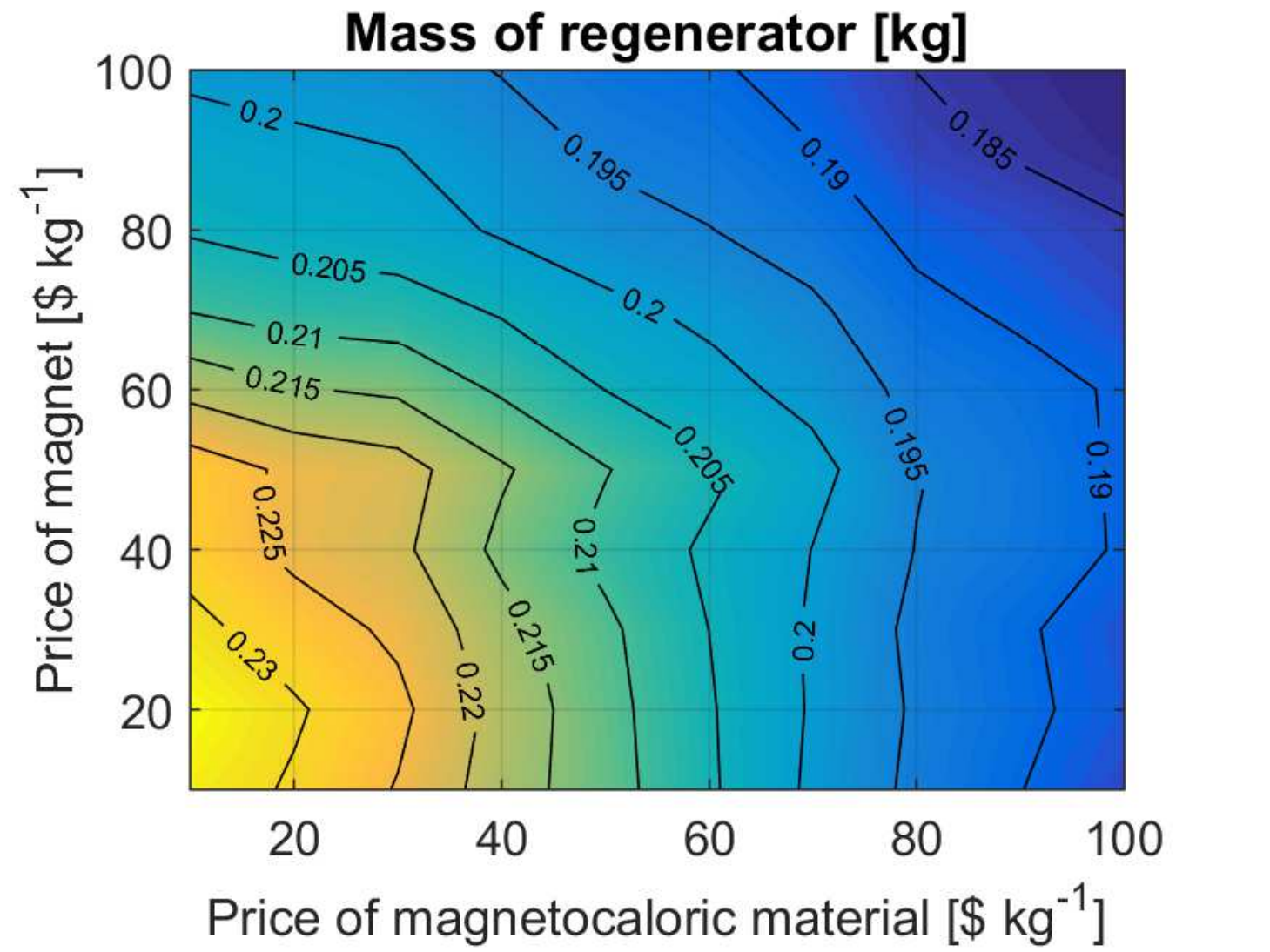}}\hspace{0.2cm}
\subfigure[50 W - 22 W]{\includegraphics[width=0.49\columnwidth]{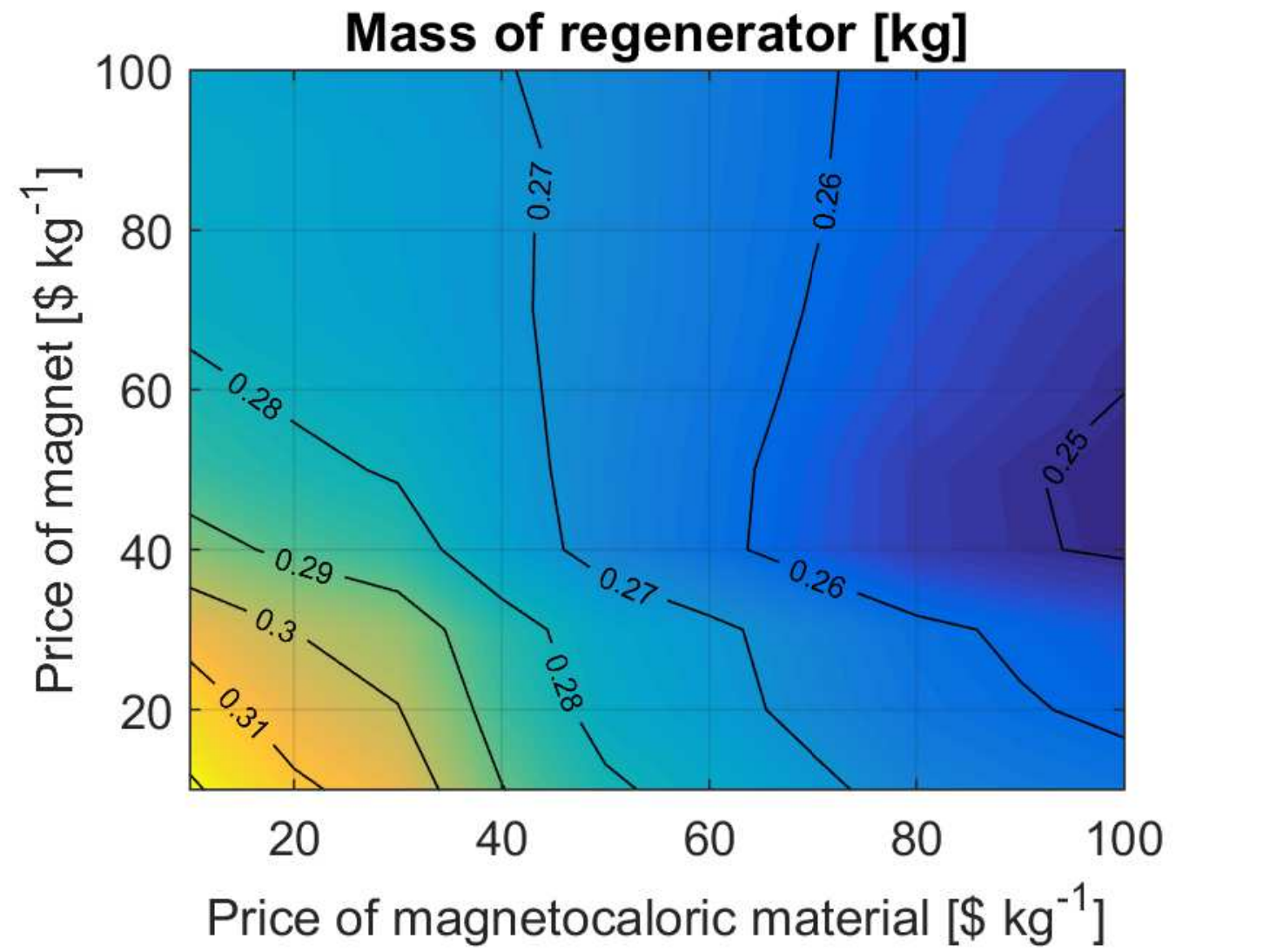}}
\caption{The mass of the regenerator.}\label{Fig_Appendix_m_reg}
\end{figure*}

\begin{figure*}[!b]
\centering
\subfigure[24.8 W]{\includegraphics[width=0.49\columnwidth]{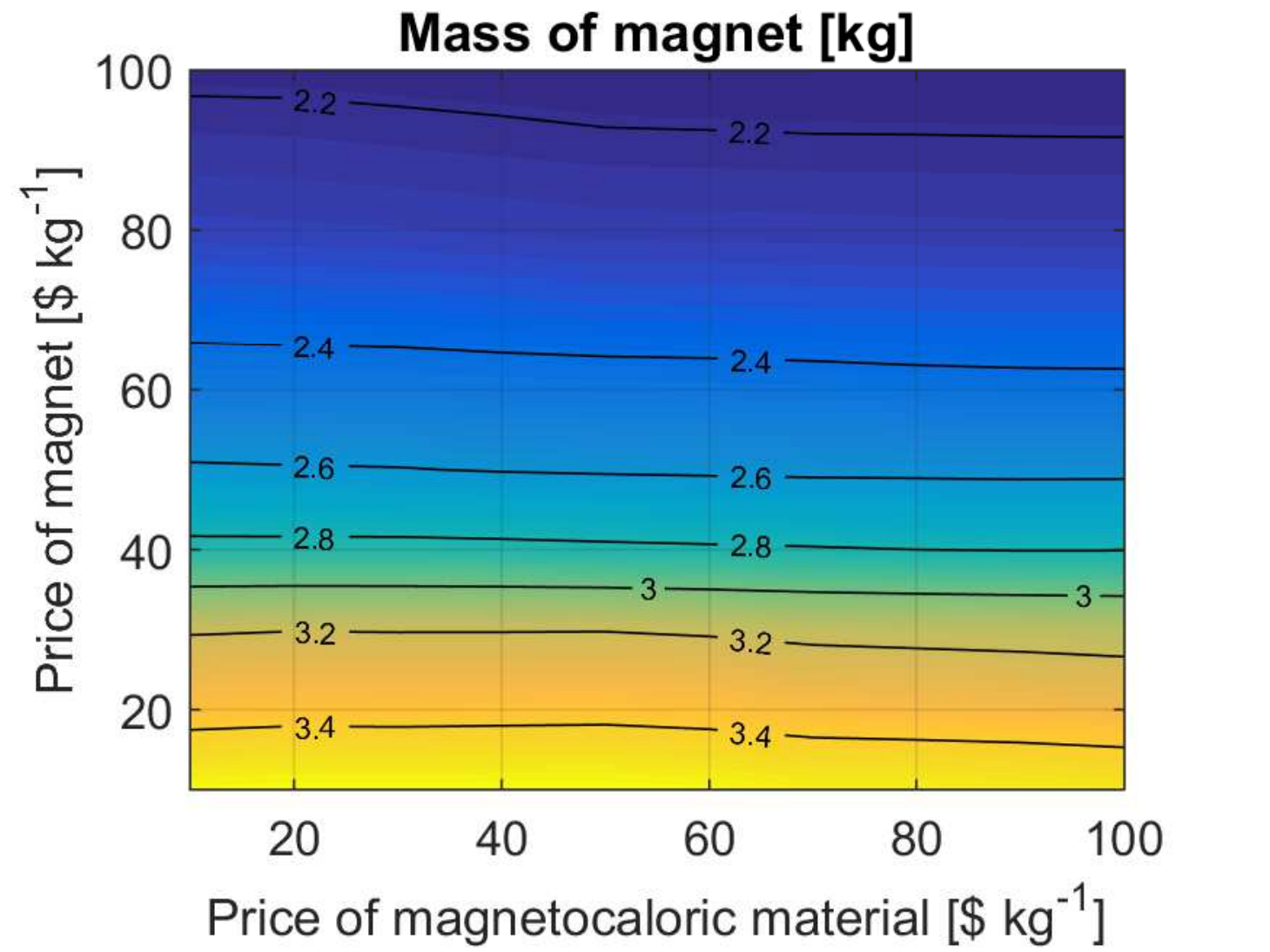}}\hspace{0.2cm}
\subfigure[50 W - 22 W]{\includegraphics[width=0.49\columnwidth]{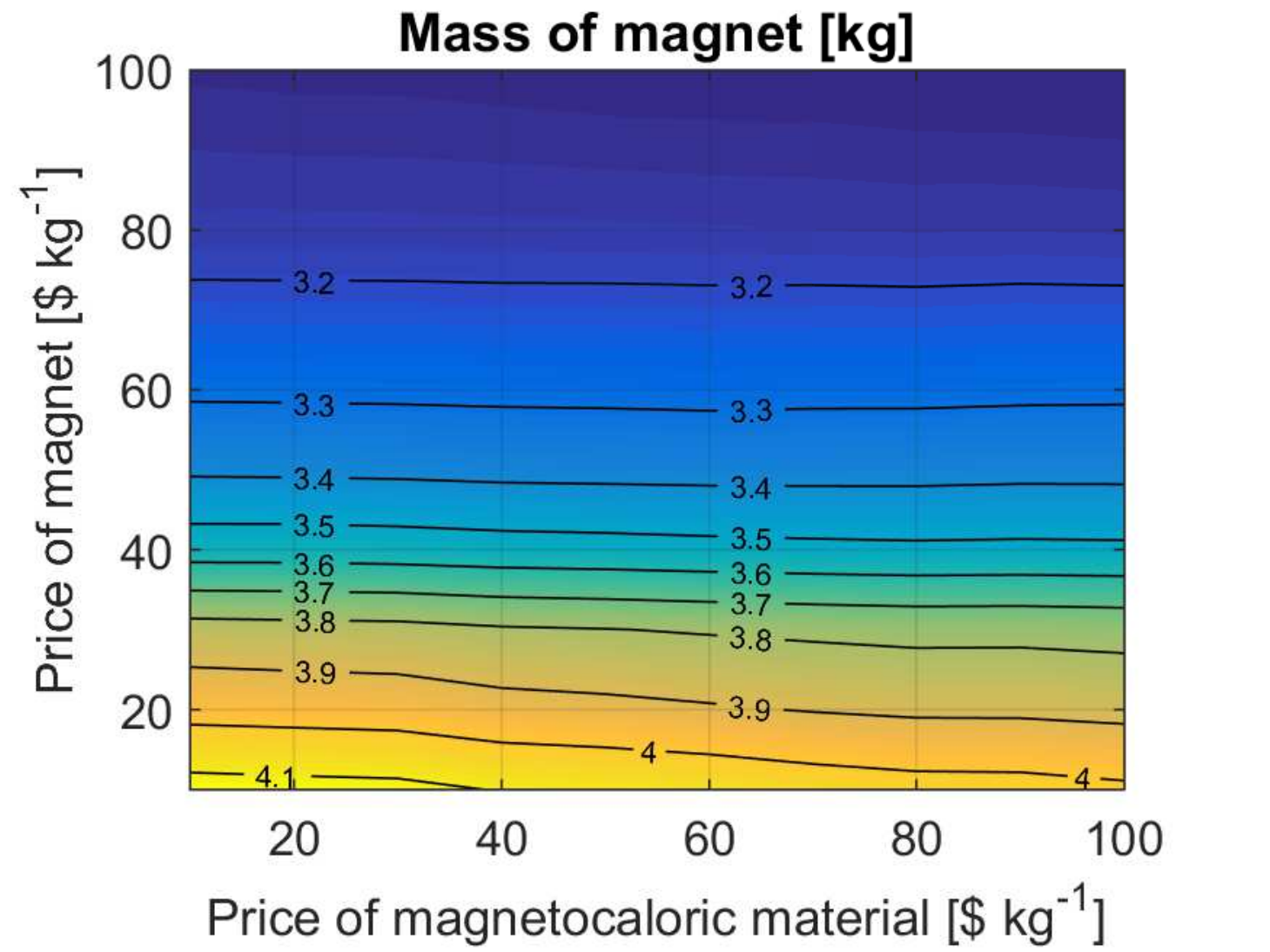}}
\caption{The mass of the magnet.}\label{Fig_Appendix_m_magnet}
\end{figure*}

\begin{figure}[!t]
\centering
\subfigure[24.8 W]{\includegraphics[width=0.49\columnwidth]{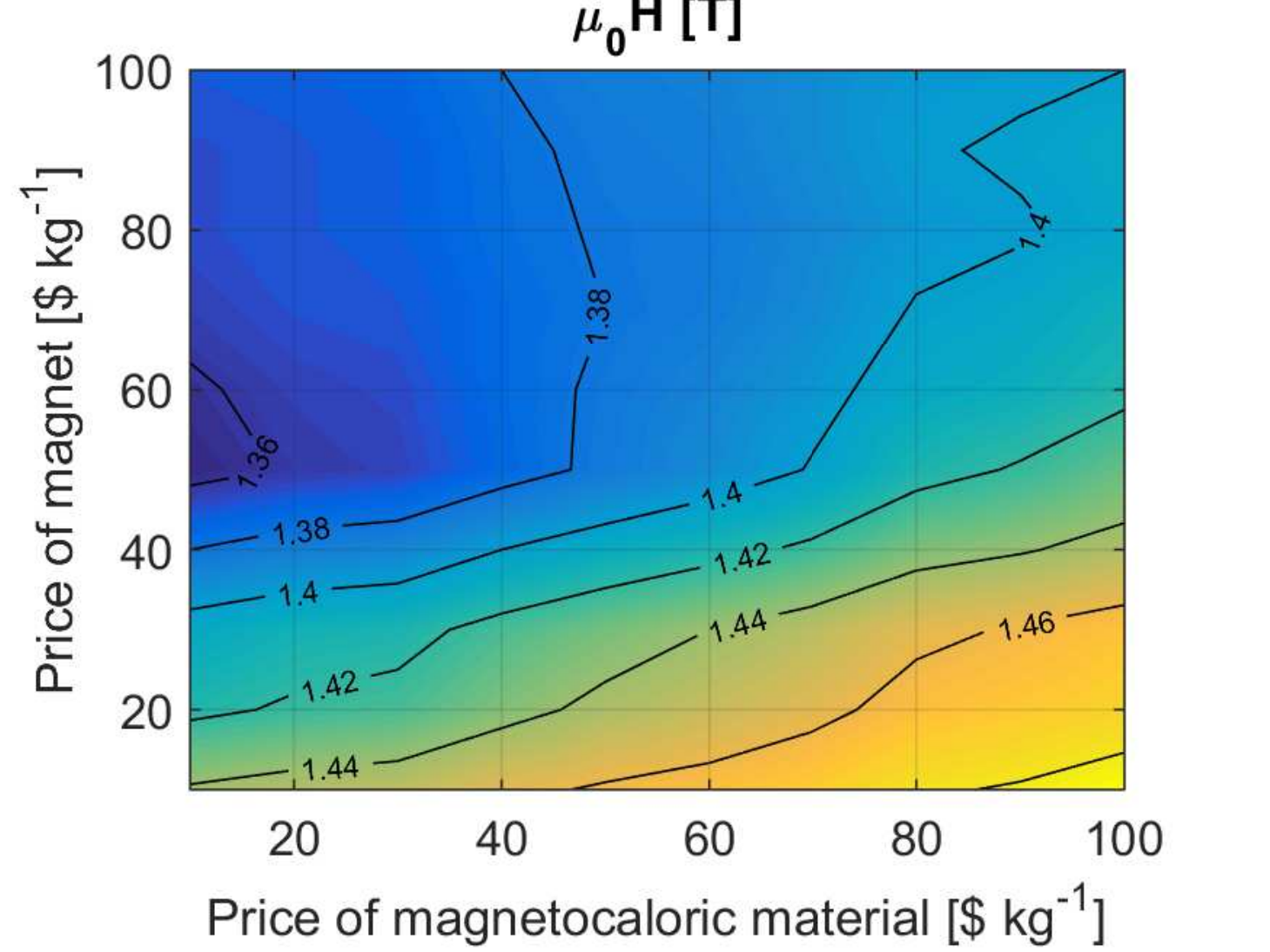}}\hspace{0.2cm}
\subfigure[50 W - 22 W]{\includegraphics[width=0.49\columnwidth]{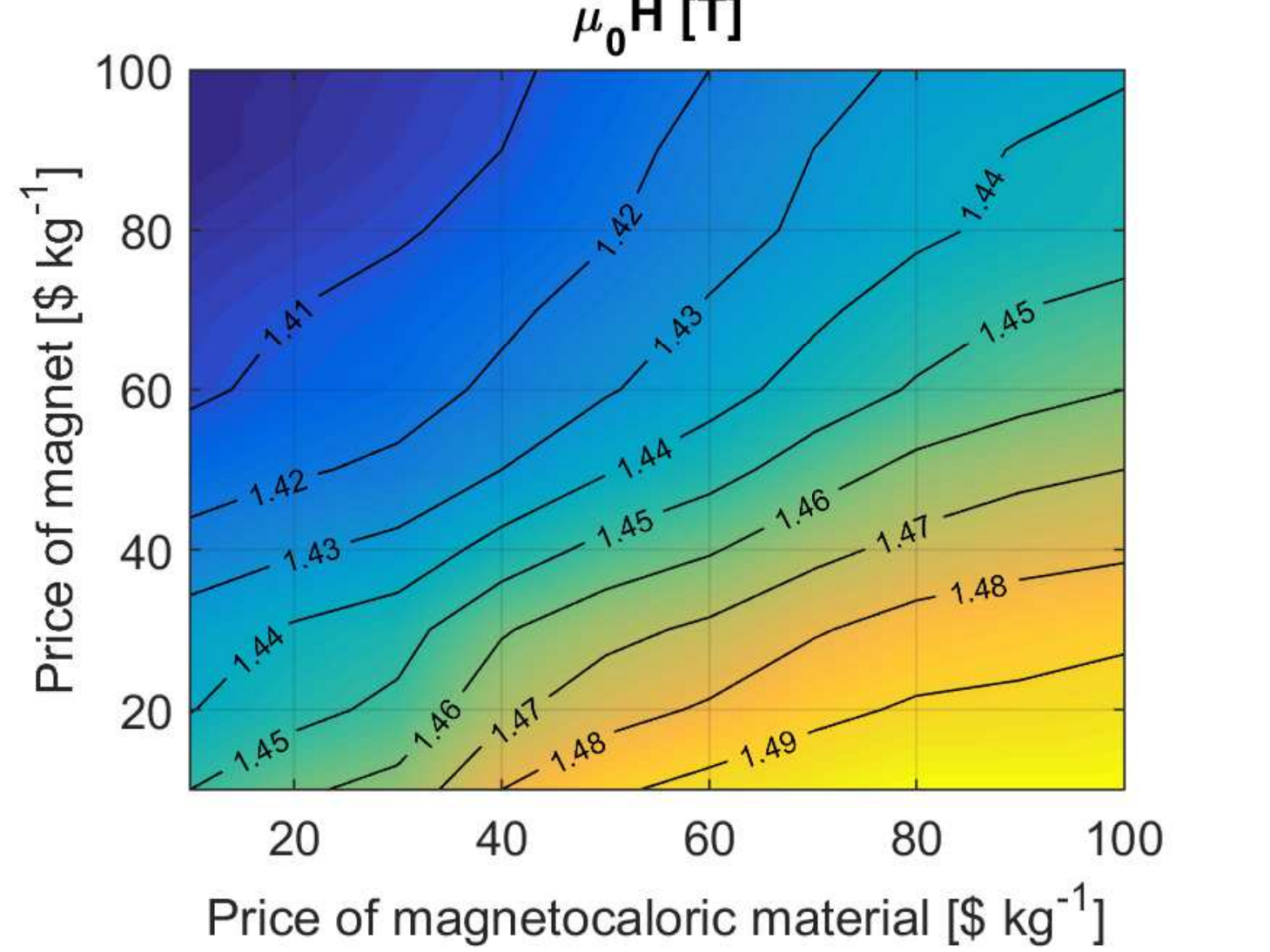}}
\caption{The magnetic field.}\label{Fig_Appendix_H}
\end{figure}

\begin{figure}[!t]
\centering
\subfigure[24.8 W]{\includegraphics[width=0.49\columnwidth]{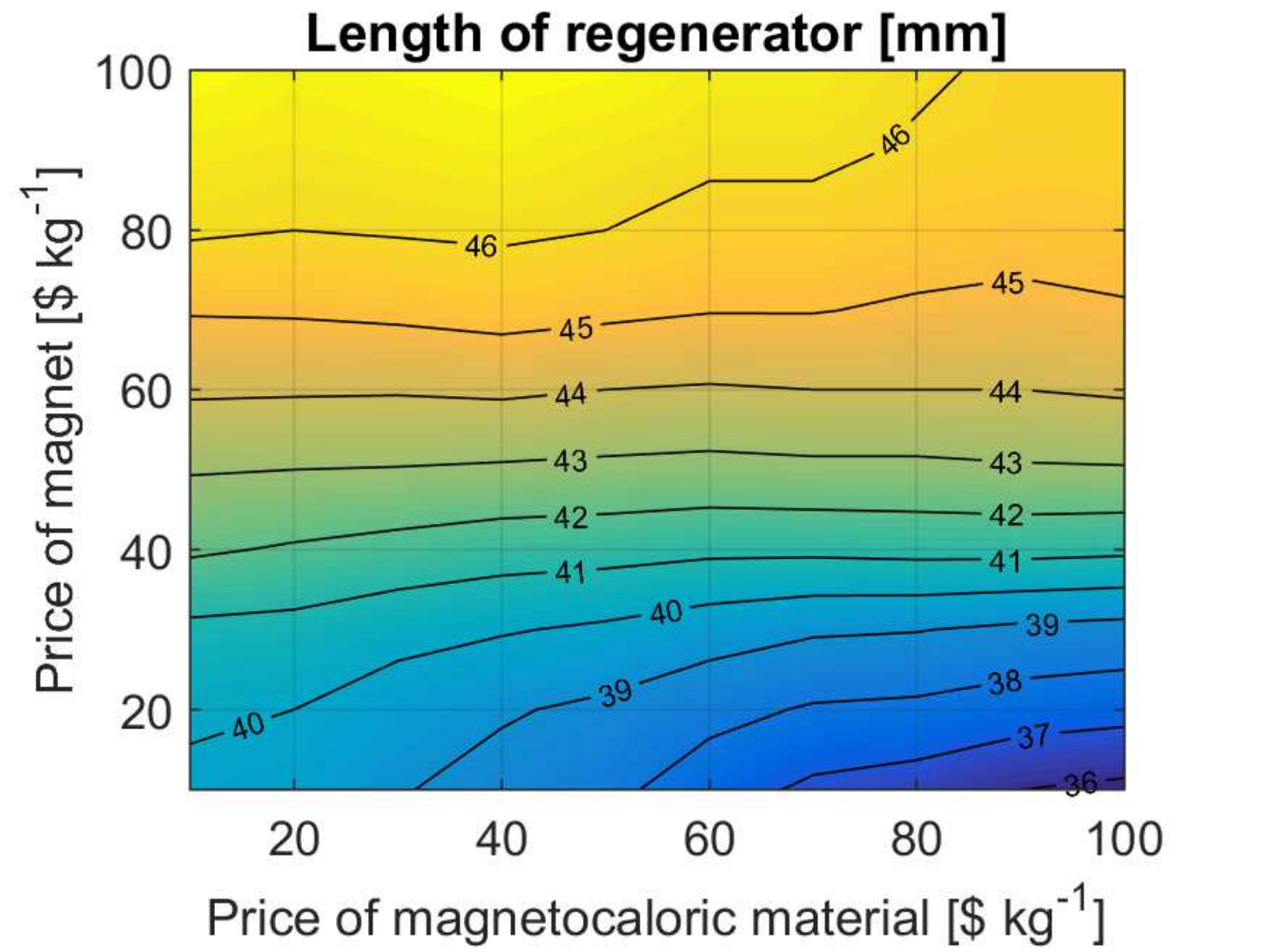}}\hspace{0.2cm}
\subfigure[50 W - 22 W]{\includegraphics[width=0.49\columnwidth]{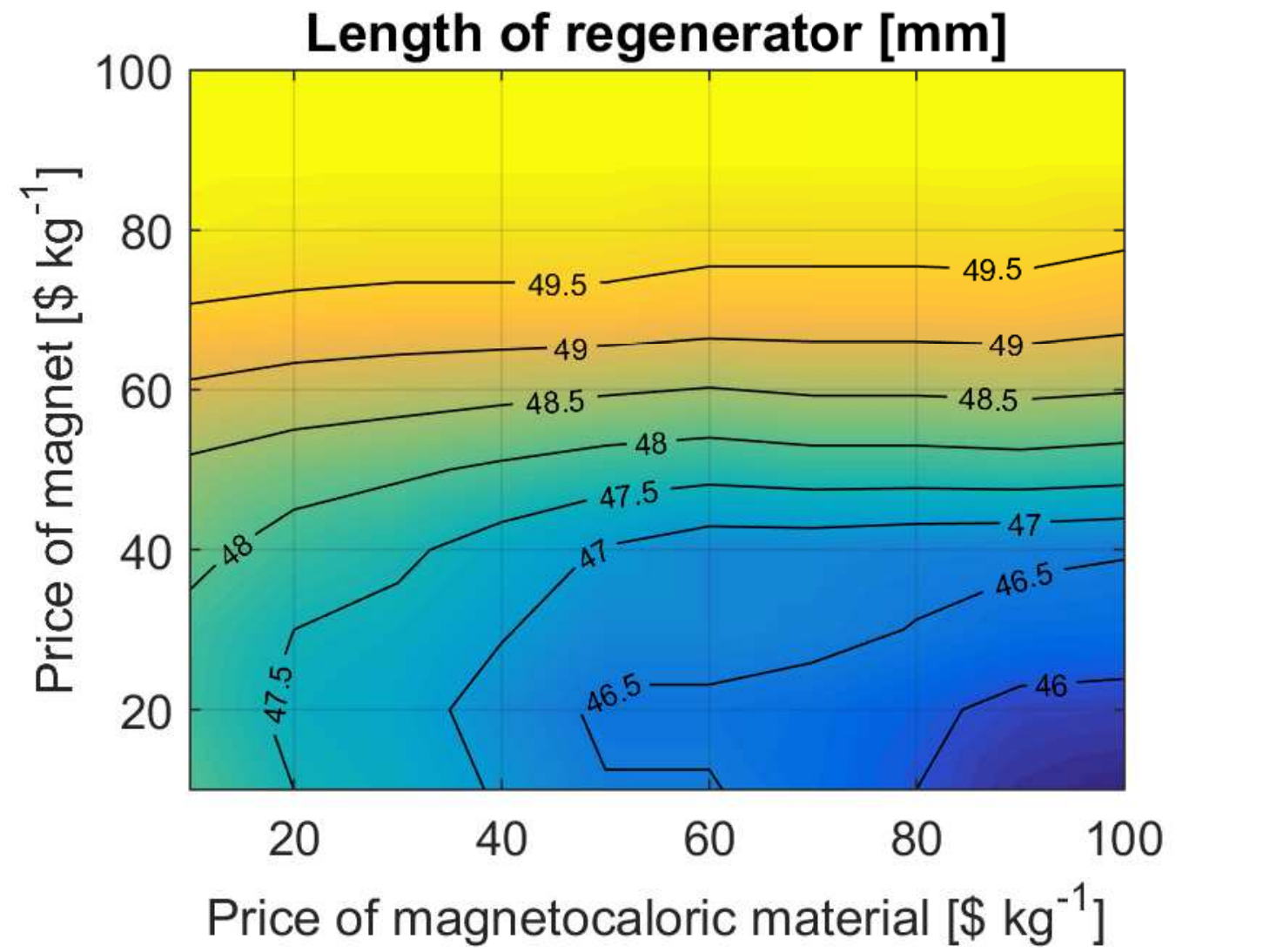}}
\caption{The length of the regenerator.}\label{Fig_Appendix_length}
\end{figure}

\begin{figure}[!t]
\centering
\subfigure[24.8 W]{\includegraphics[width=0.49\columnwidth]{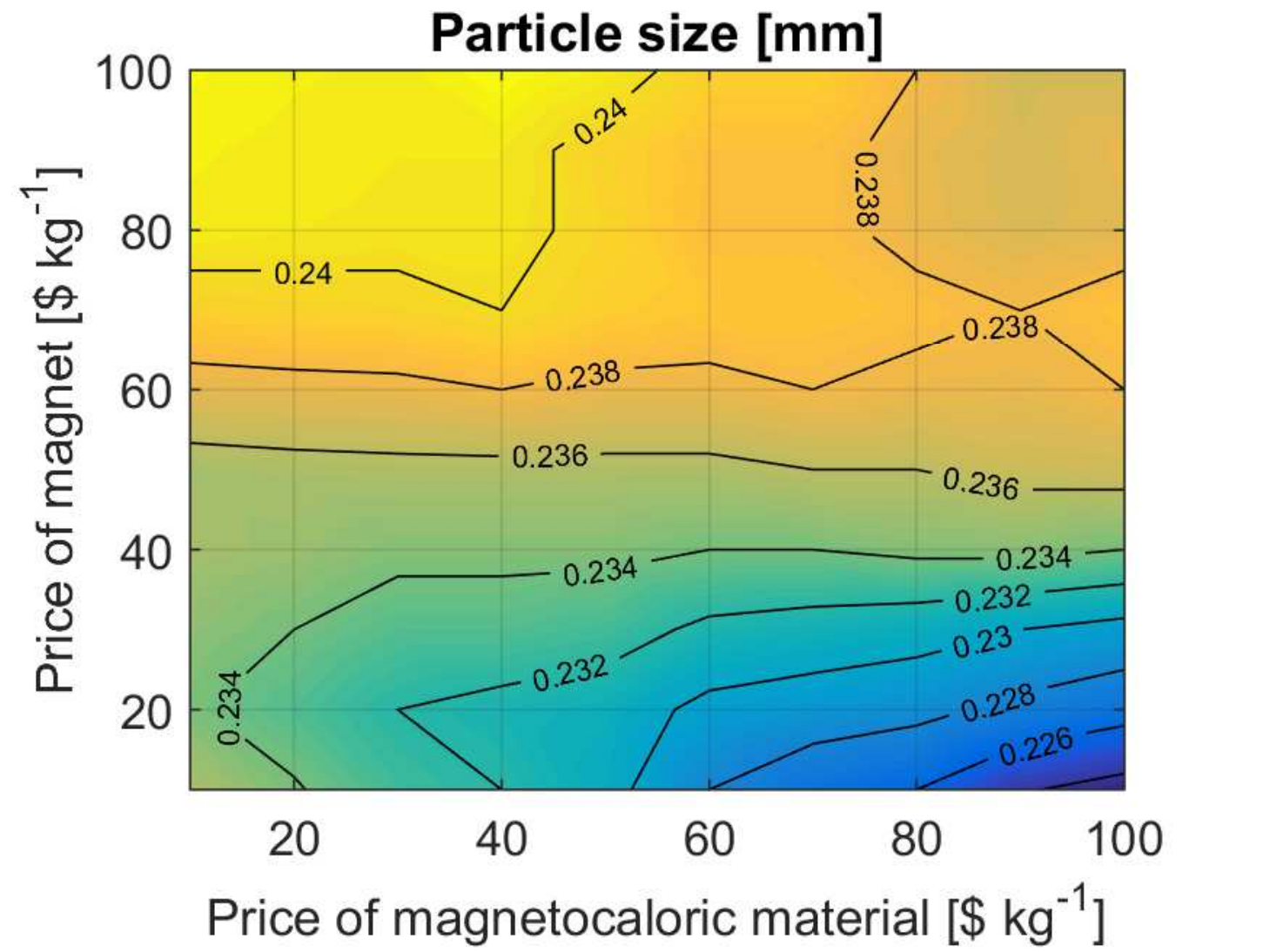}}\hspace{0.2cm}
\subfigure[50 W - 22 W]{\includegraphics[width=0.49\columnwidth]{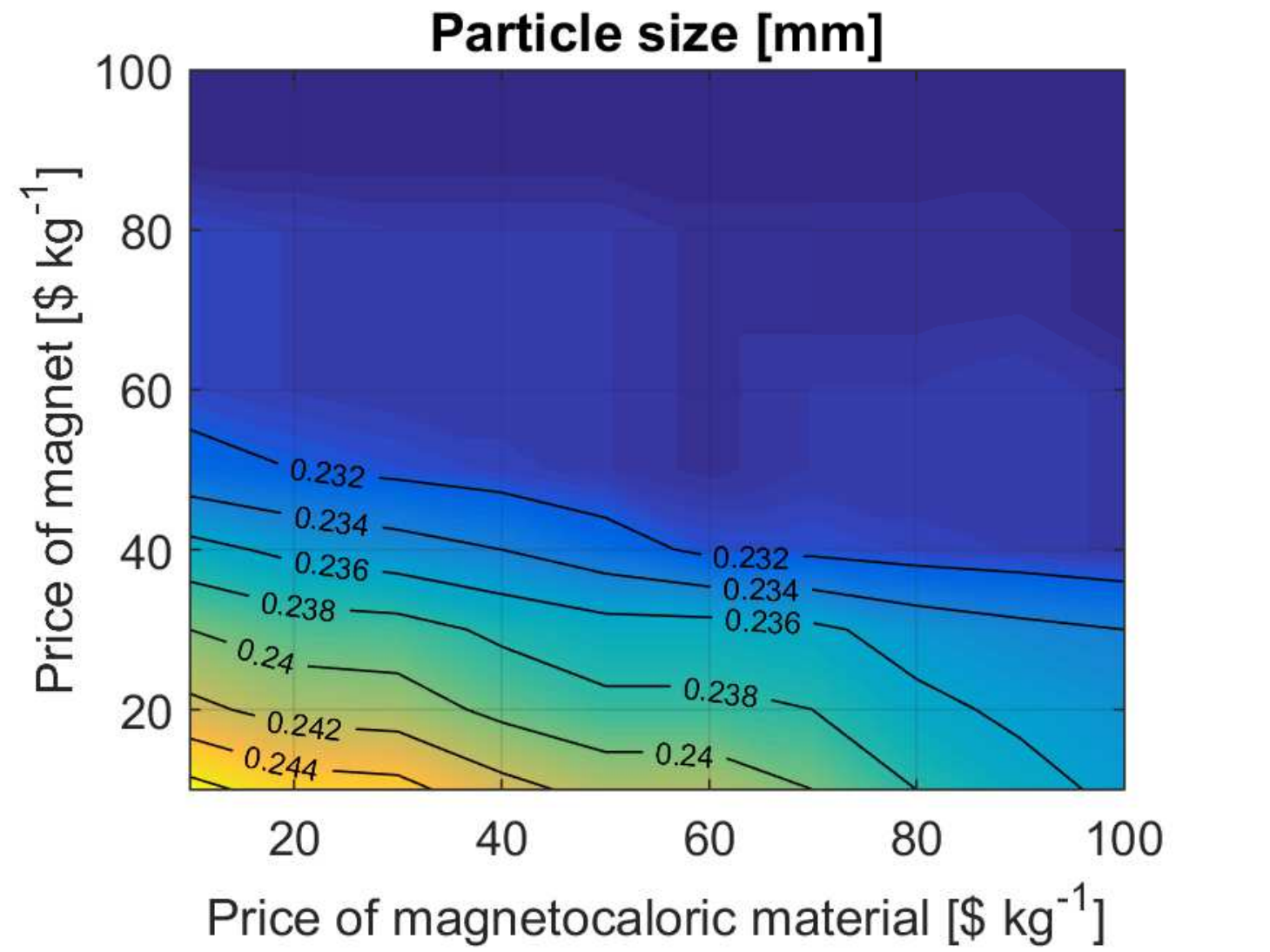}}
\caption{The particle size.}\label{Fig_Appendix_particle_size}
\end{figure}

\begin{figure}[!t]
\centering
\subfigure[24.8 W]{\includegraphics[width=0.49\columnwidth]{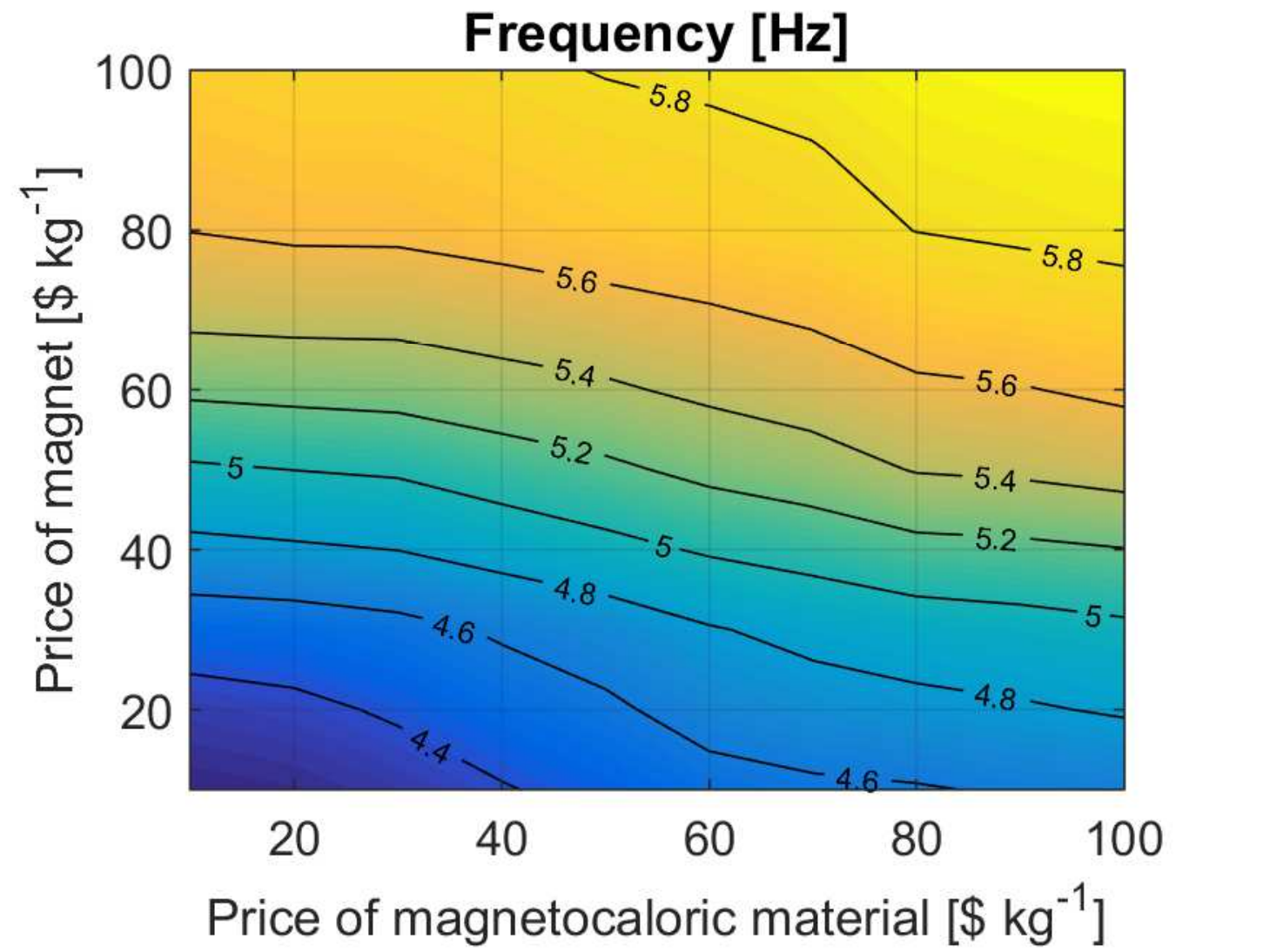}}\hspace{0.2cm}
\subfigure[50 W - 22 W @ 50 W]{\includegraphics[width=0.49\columnwidth]{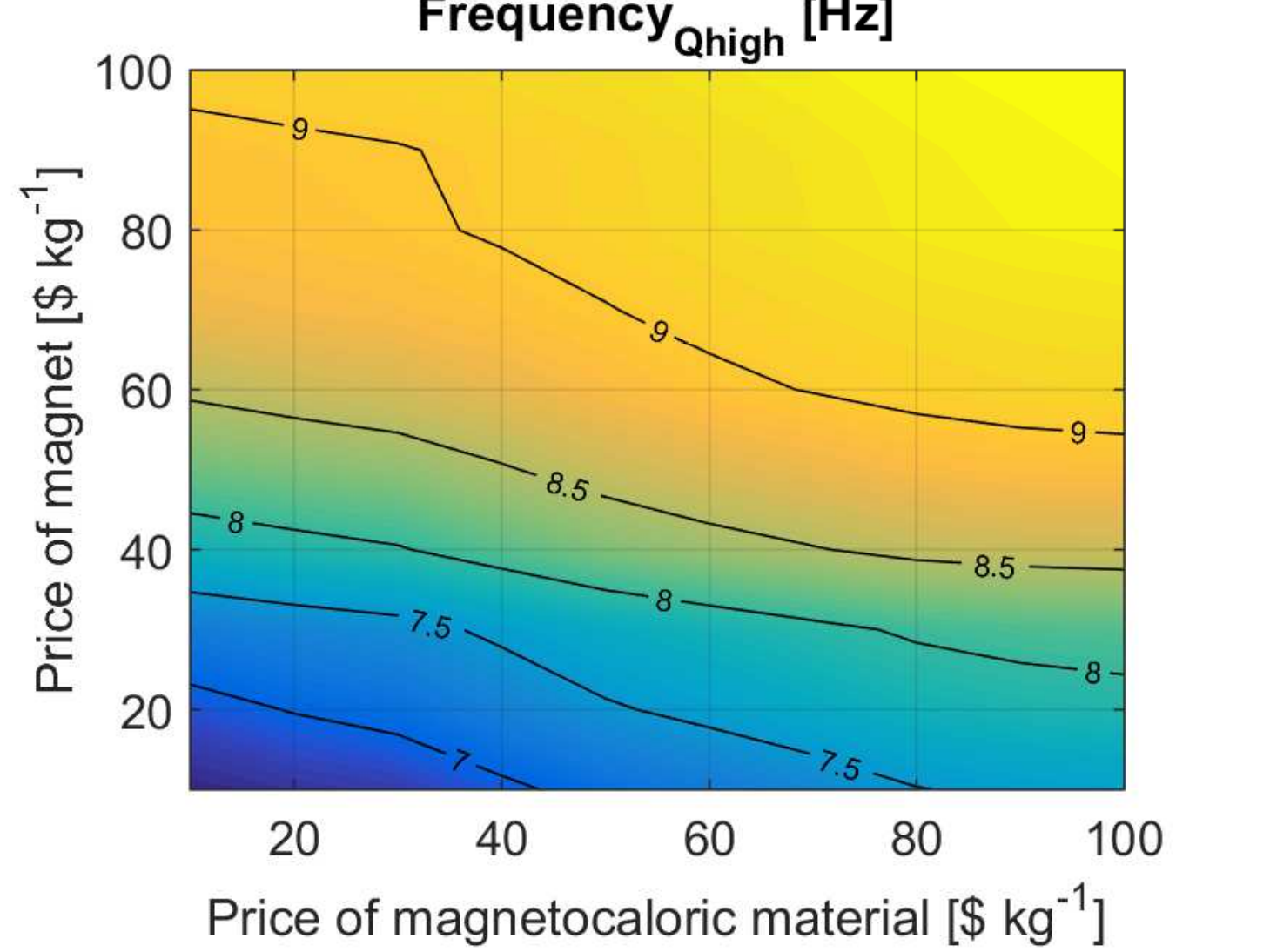}}
\subfigure[50 W - 22 W @ 22 W]{\includegraphics[width=0.49\columnwidth]{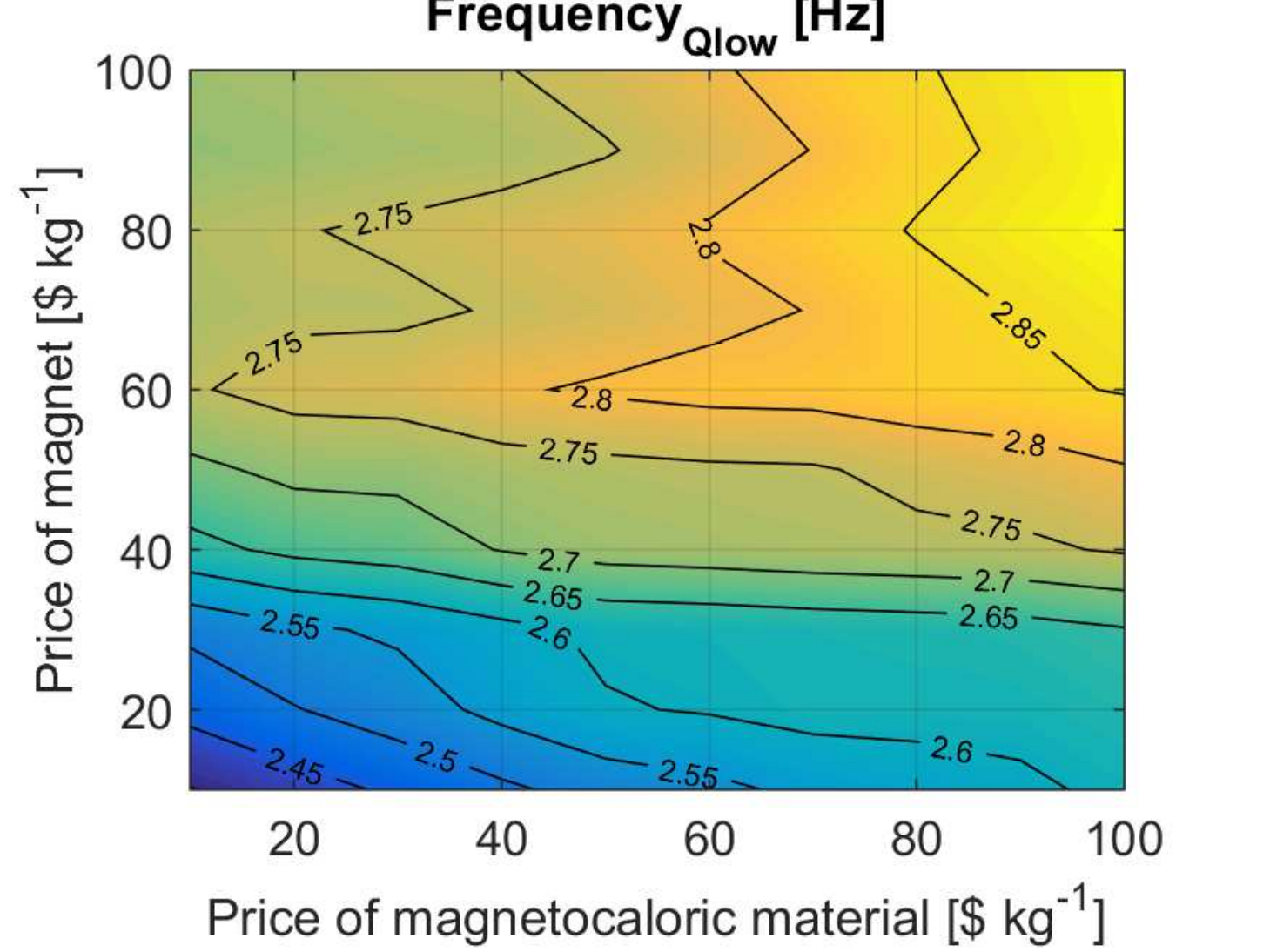}}
\caption{The frequency.}\label{Fig_Appendix_frequency}
\end{figure}

\begin{figure}[!t]
\centering
\subfigure[24.8 W]{\includegraphics[width=0.49\columnwidth]{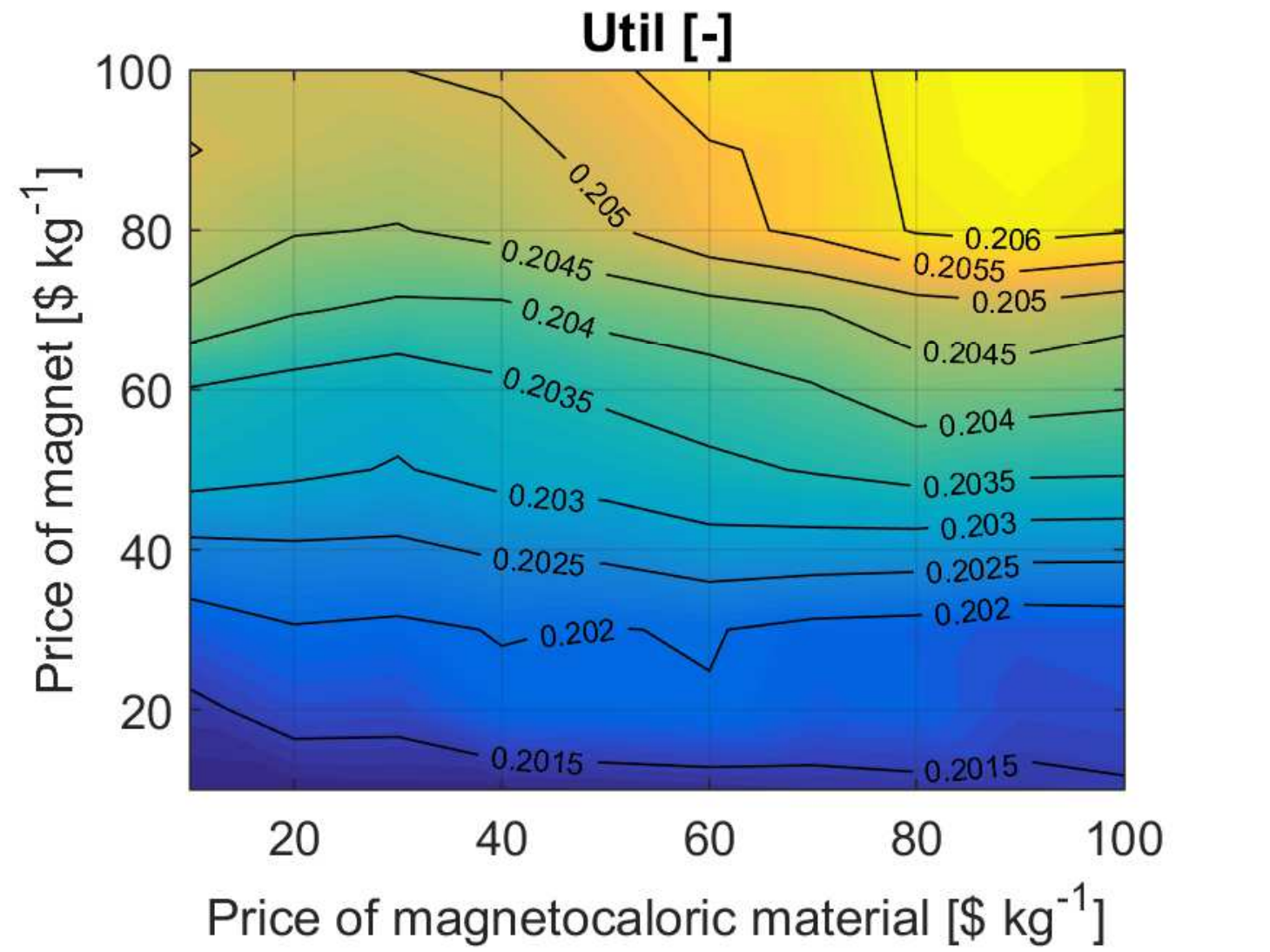}}\hspace{0.2cm}
\subfigure[50 W - 22 W @ 50 W]{\includegraphics[width=0.49\columnwidth]{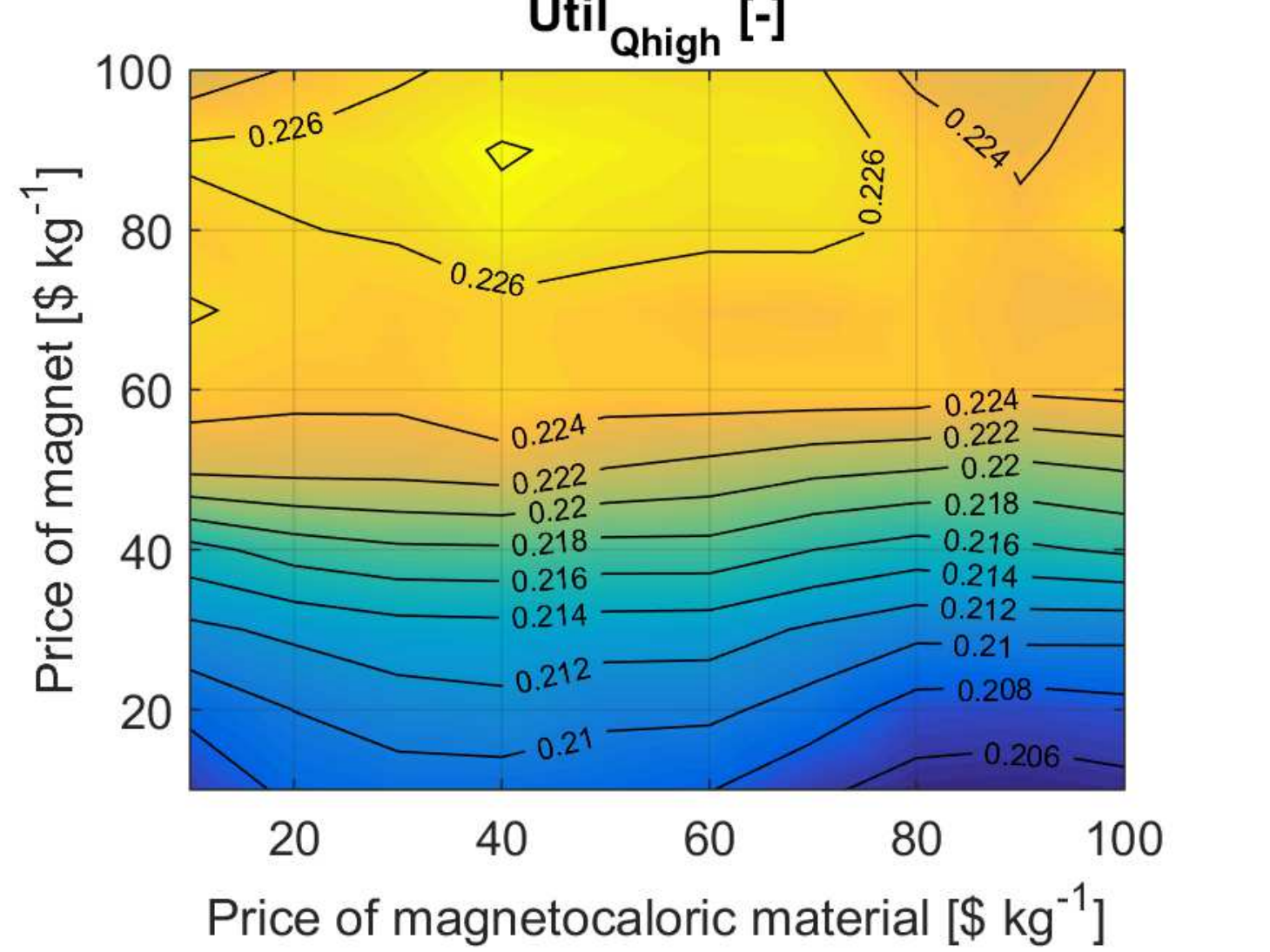}}
\subfigure[50 W - 22 W @ 22 W]{\includegraphics[width=0.49\columnwidth]{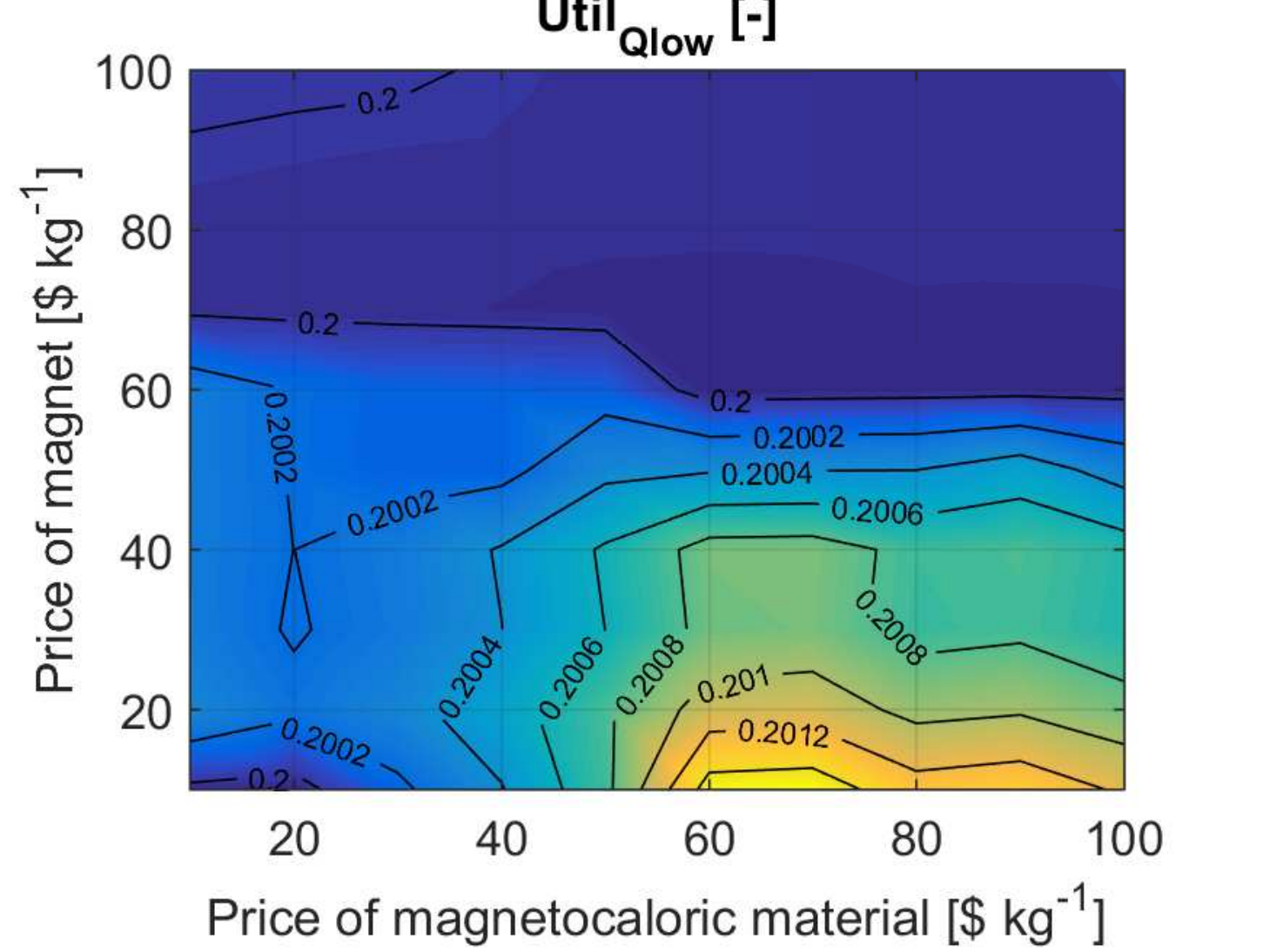}}
\caption{The utilization.}\label{Fig_Appendix_utilization}
\end{figure}

\begin{figure}[!t]
\centering
\subfigure[24.8 W]{\includegraphics[width=0.49\columnwidth]{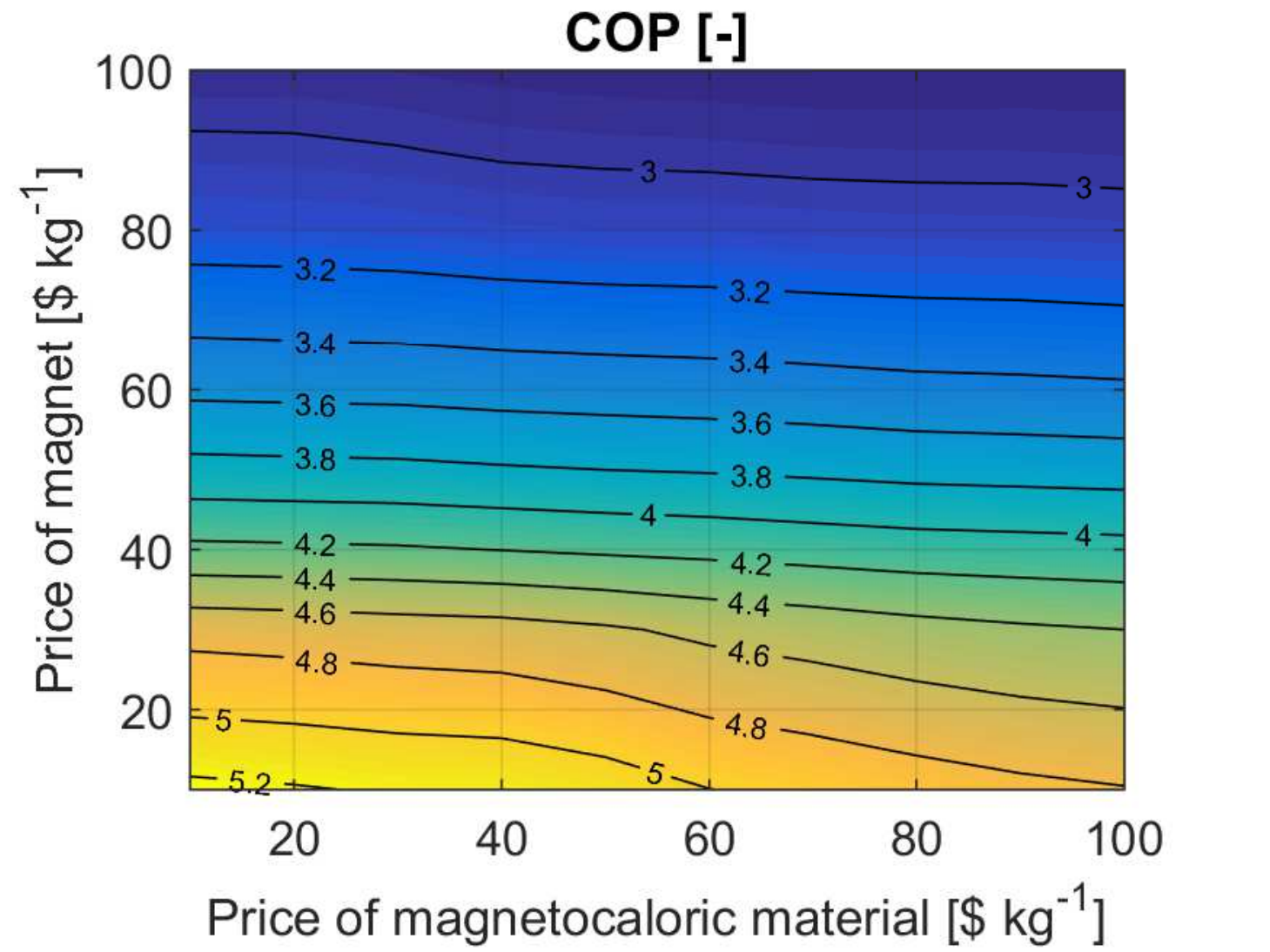}}\hspace{0.2cm}
\subfigure[50 W - 22 W Average]{\includegraphics[width=0.49\columnwidth]{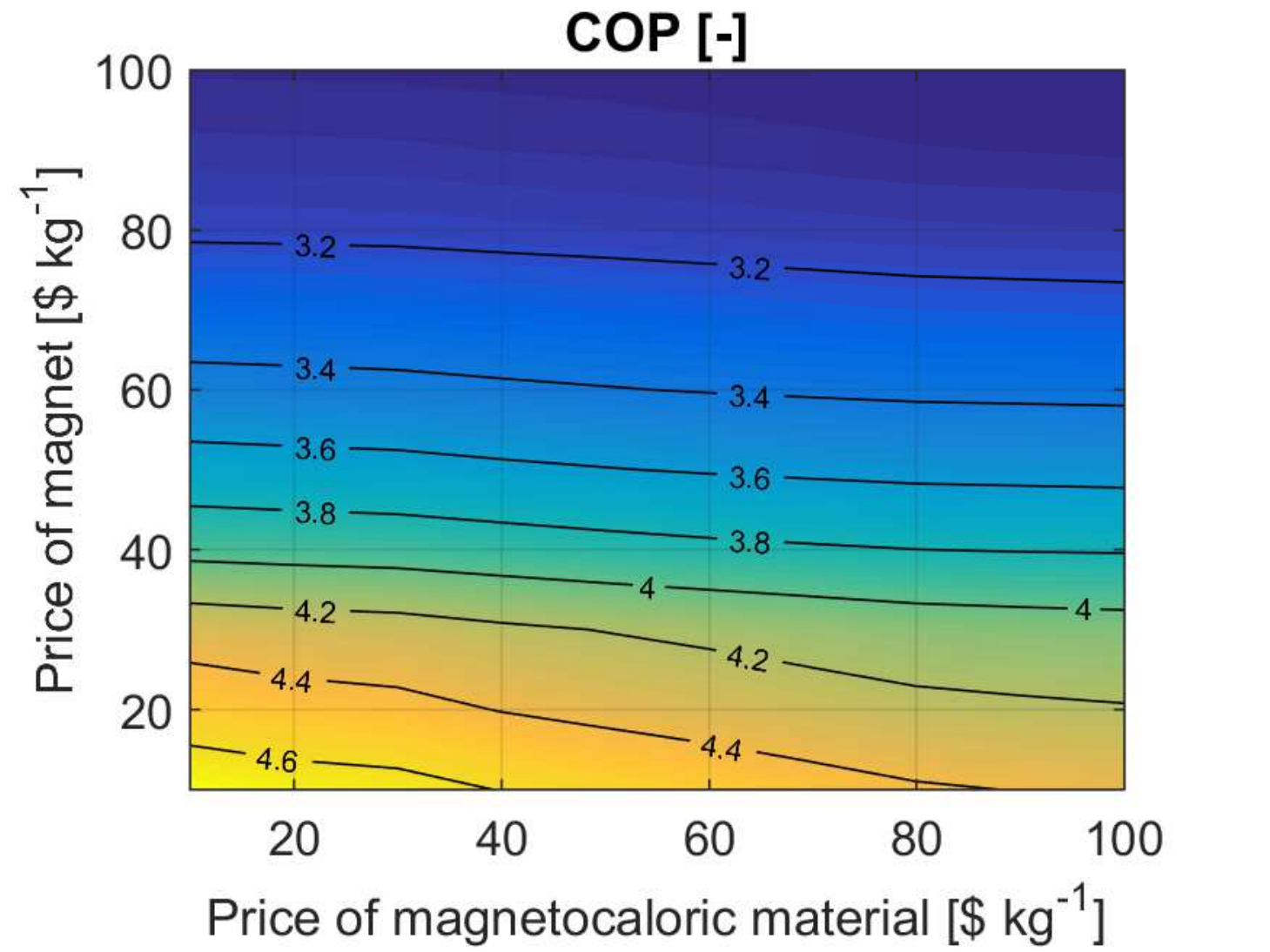}}
\subfigure[50 W - 22 W @ 50 W]{\includegraphics[width=0.49\columnwidth]{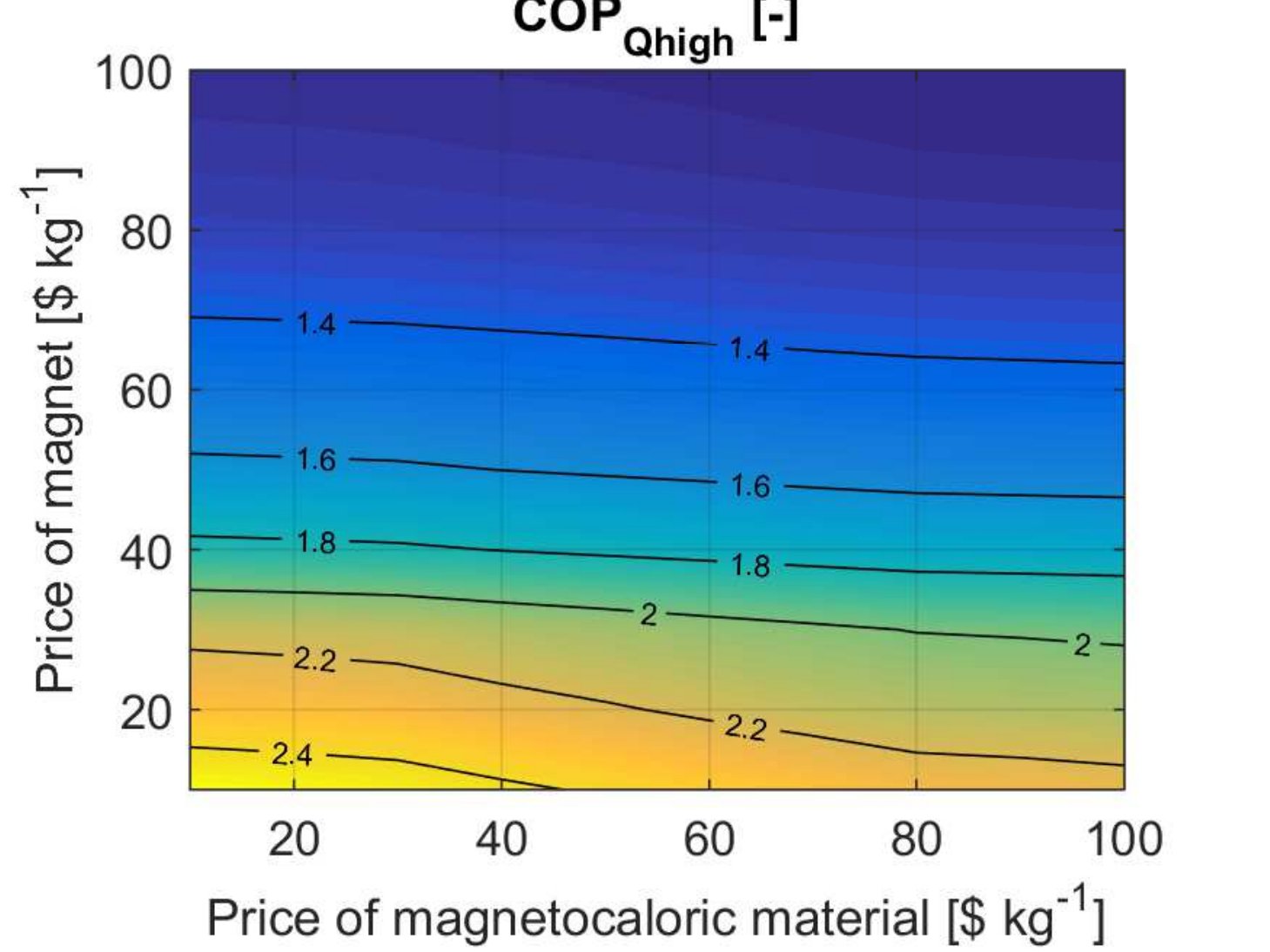}}
\subfigure[50 W - 22 W @ 22 W]{\includegraphics[width=0.49\columnwidth]{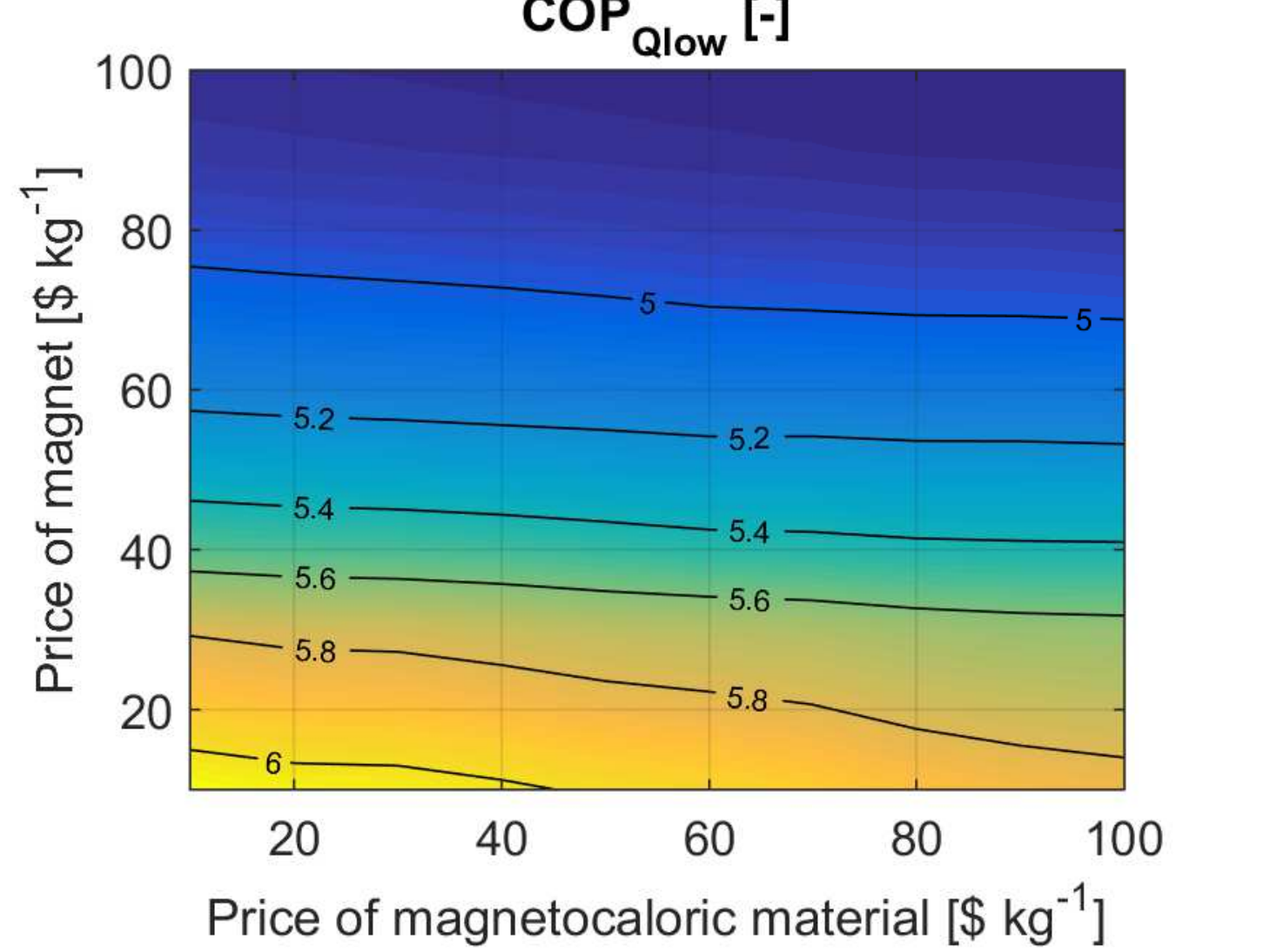}}
\caption{The COP.}\label{Fig_Appendix_COP}
\end{figure}

In the figures below are given the operation parameters for a refrigerator cooling 24.8 W at all times and for a system cooling 50 W for 10 \% of the time and 22 W for the remaining 90 \% of the time. Both have an expected lifetime of 15 years and the operating parameters are shown as a function of the price of the magnet material and the magnetocaloric material. The figures show the mass of the regenerator (Fig. \ref{Fig_Appendix_m_reg}), the mass of the magnet (Fig. \ref{Fig_Appendix_m_magnet}), the magnetic field (Fig. \ref{Fig_Appendix_H}), the length of the regenerator (Fig. \ref{Fig_Appendix_length}), the particle size (Fig. \ref{Fig_Appendix_particle_size}), the frequency (Fig. \ref{Fig_Appendix_frequency}), the
utilization (Fig. \ref{Fig_Appendix_utilization}) and the COP (Fig. \ref{Fig_Appendix_COP}).

\clearpage

\twocolumn

\end{document}